\documentclass{webofc}
\usepackage[varg]{txfonts}   
\pdfoutput=1 
\begin{document}
\title{Introduction to optical/IR interferometry: history
and basic principles}
%
%

\author{\firstname{Jean} \lastname{Surdej}\inst{1}\fnsep\thanks{\email{jsurdej@ulg.ac.be}} 
}

\institute{Institute of Astrophysics and Geophysics, Li\`{e}ge University, All\'ee du 6 Ao\^{u}t 19c, 4000 Li\`{e}ge,
Belgium 
          }

\abstract{%
 The present notes refer to a 3-hour lecture delivered on 27
September 2017 in Roscoff during the 2017 Evry Schatzman School. It
concerns a general introduction to optical/IR interferometry, including a
brief history, a presentation of the basic principles, some important theorems
and relevant applications. The layout of these lecture notes is as follows. After a short
introduction, we proceed with some reminders concerning the
representation of a field of electromagnetic radiation. We then present a
short history of interferometry, from the  first experiment of Fizeau and Stefan 
to modern optical interferometers. We then discuss the notions of light
coherence, including the theorem of Zernicke-van Cittert and describe the
principle of interferometry using two telescopes. We present some
examples of modern interferometers and typical results obtained with
these. Finally, we address three important theorems: the fundamental
theorem, the convolution theorem and the Wiener-Khinchin theorem which
enable to get a better insight into the field of optical/IR interferometry.
}
\maketitle
\section{Introduction}
\label{intro}
\begin{figure}[h]
\sidecaption
\centering
\includegraphics[width=9cm]{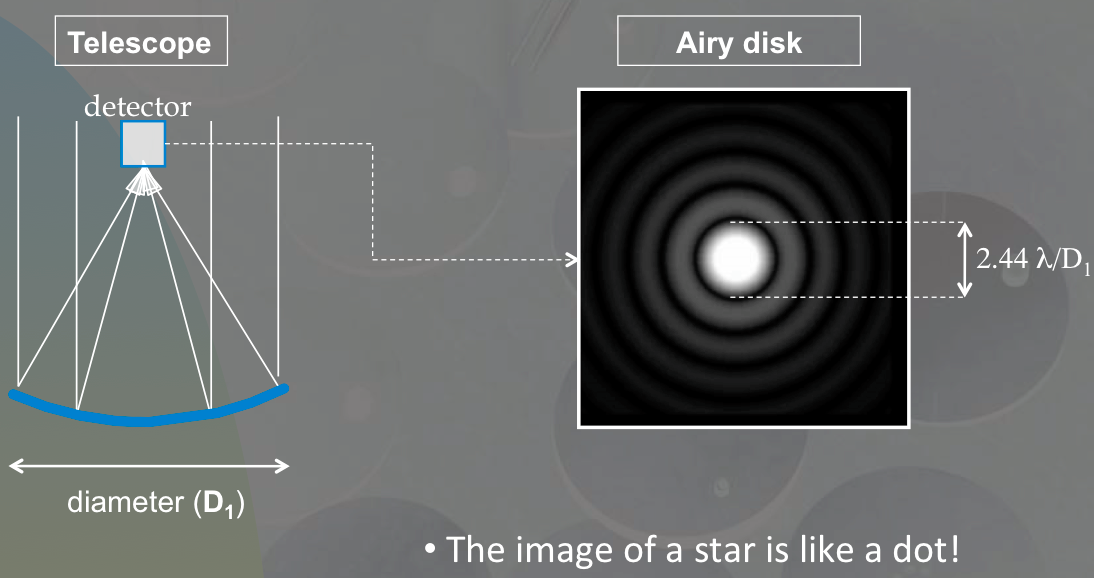}
\caption{Airy disk of a point-like star recorded in the focal plane of a telescope with diameter $D_1$. The angular diameter of the Airy disk is 2.44 $\lambda / D_1$.}
\label{Fig_1_final}       
\end{figure}

\begin{figure}[h]
\sidecaption
\centering
\includegraphics[width=9cm]{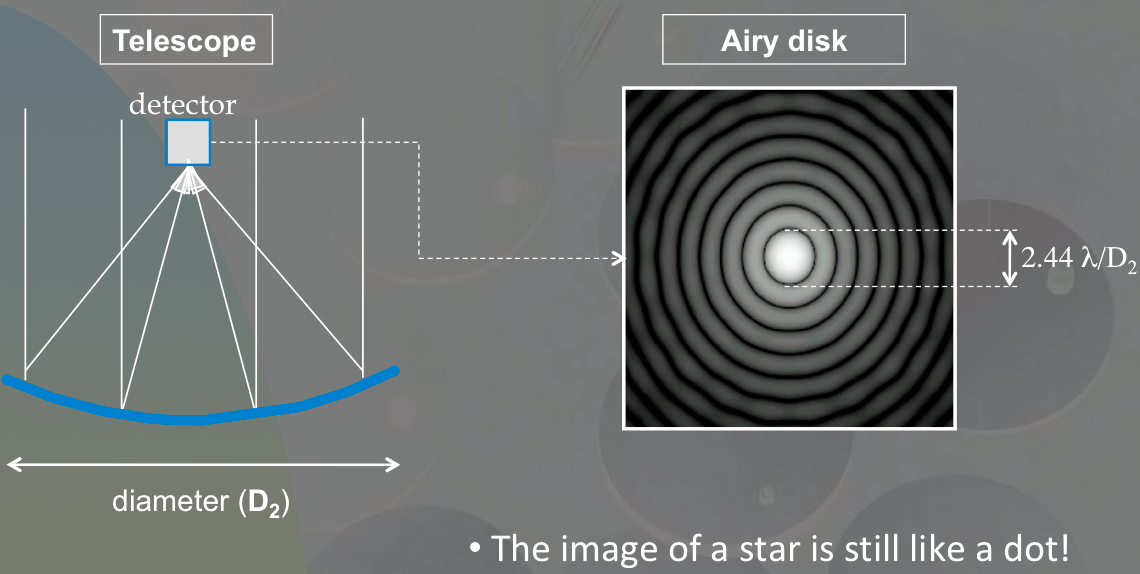}
\caption{As the diameter of a telescope increases ($D_2 > D_1$), the Airy disk of a point-like star gets smaller. }
\label{Fig_2_final}       
\end{figure}

\begin{figure}[h]
\sidecaption
\centering
\includegraphics[width=9cm]{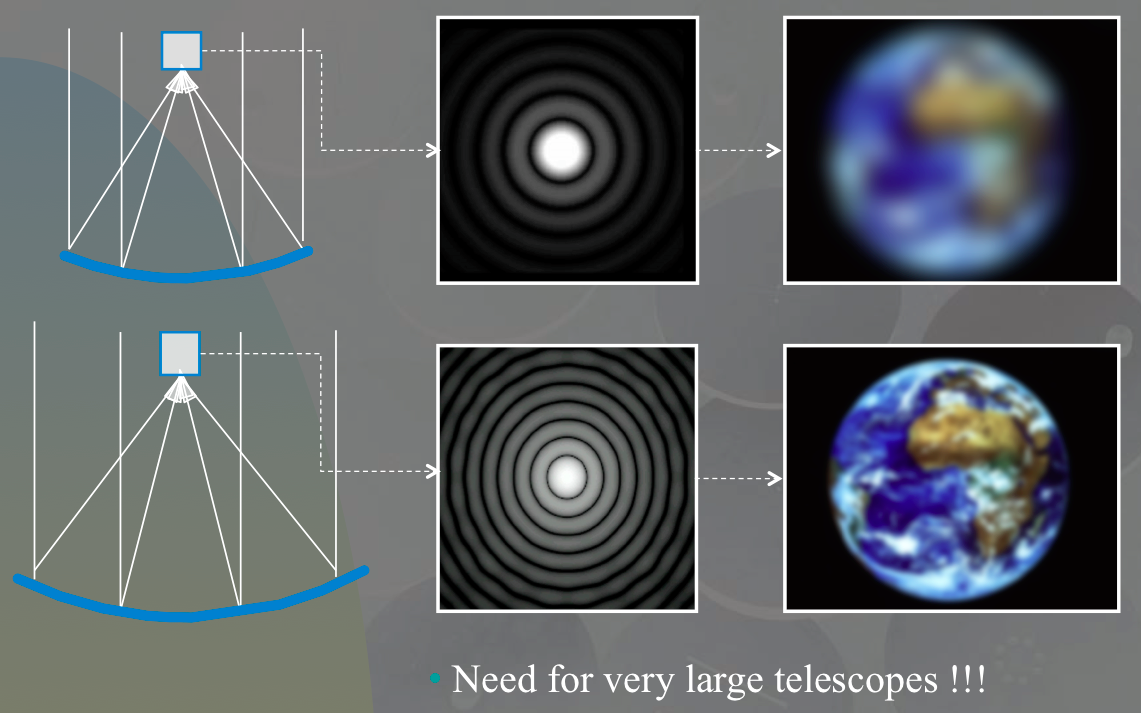}
\caption{While observing an extended celestial object (cf. an Earth-like planet) above the atmosphere, we see more details as the diameter of the telescope increases.  }
\label{Fig_3_final}       
\end{figure}

\begin{figure}[h]
\sidecaption
\centering
\includegraphics[width=9cm]{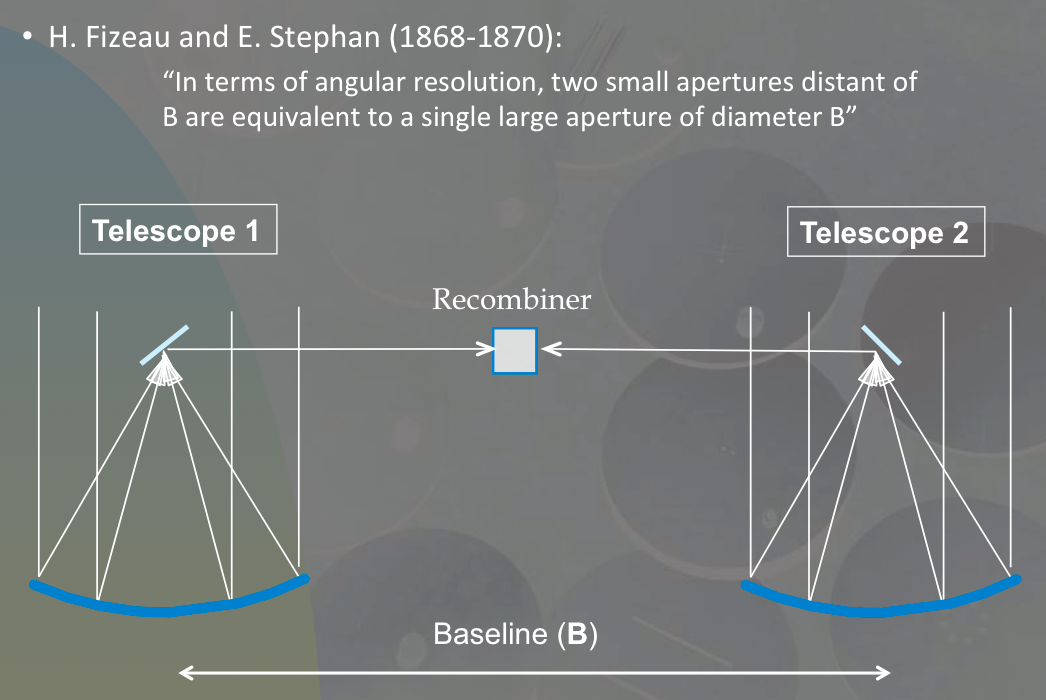}
\caption{Fizeau and Stephan proposed to recombine the light from two independent telescopes separated by a baseline $B$ to recover the same angular resolution as that given by a single dish telescope having a diameter $B$.  }
\label{Fig_4_final}       
\end{figure}

\begin{figure}[h]
\sidecaption
\centering
\includegraphics[width=9cm]{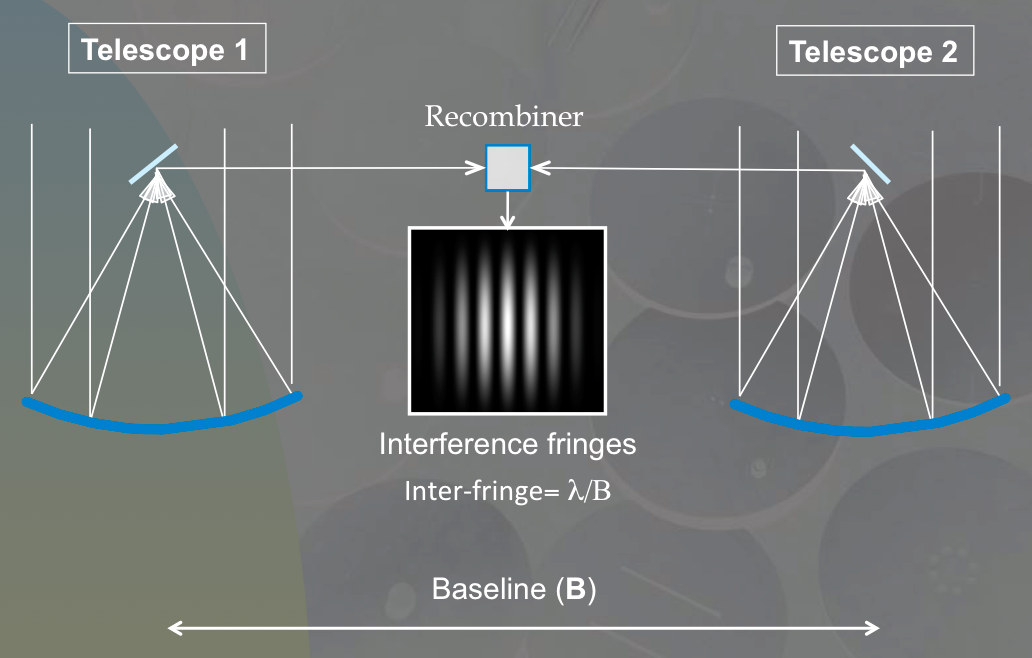}
\caption{When recombining the monochromatic light of two independent telescopes, there results the formation of a pattern of bright and dark fringes superimposed over the combined Airy disk. The angular inter-fringe separation is equal to $\lambda / B$. }
\label{Fig_5_final}       
\end{figure}

\begin{figure}[h]
\sidecaption
\centering
\includegraphics[width=9cm]{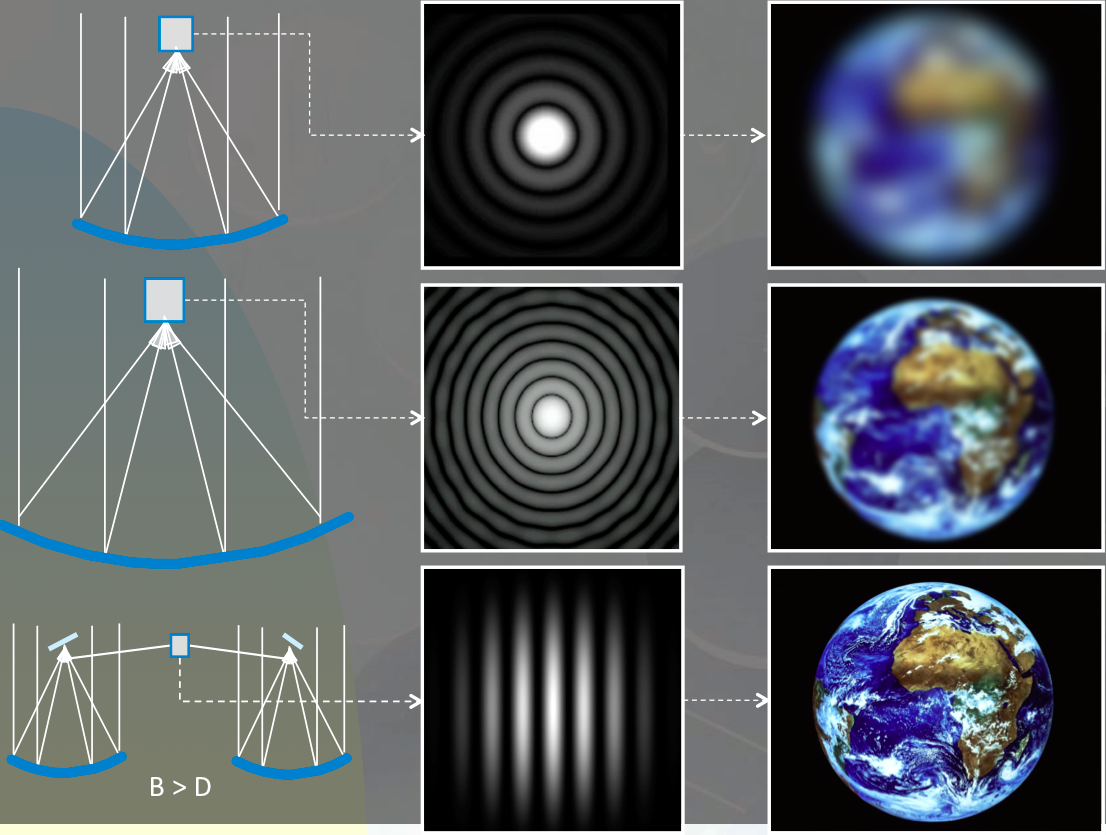}
\caption{Improvement expected in angular resolution while observing an extended celestial source (cf. an Earth-like planet) with telescopes of increasing size ($D_2 > D_1$) and with an interferometer composed of two telescopes separated by a baseline $B > D$. }
\label{Fig_6_final}       
\end{figure}

In the absence of the Earth atmosphere above a ground-based telescope equipped with a mirror having a diameter $D_1$, Figure~\ref{Fig_1_final} illustrates the image one would observe from a point-like star recorded in the focal plane in monochromatic light at a wavelength $\lambda$. It is a dot of light, the well-known Airy disk, which angular radius measured in radian is simply given by 1.22 $ \lambda / D_1$. Unfortunately, the Airy disk does not contain any information relative to the star being imaged, irrespective of its size, shape, effective temperature, luminosity, distance, etc. A larger telescope with a diameter $D_2 > D_1$, would similarly lead to a smaller Airy disk (1.22 $\lambda / D_2$) of light for the star being imaged (see Fig.~\ref{Fig_2_final}), providing a slightly better angular resolution image but with no more specific information related to the star. While observing an extended celestial source (cf. a distant resolved Earth-like planet as shown in Fig.~\ref{Fig_3_final}), more details are seen with the telescope having a larger diameter. The dream of astronomers is therefore to construct always larger telescopes but presently there is a limit (D $\sim$ 40m) over which it is technologically difficult to construct a single mirror telescope (cf. the ELT, TMT, GMT projects).  \\
Fortunately, in 1868 Fizeau and Stephan just realized that ''In terms of angular resolution, two small apertures distant of $B$ are equivalent to a single large aperture of diameter $B$.'' (see Fig.~\ref{Fig_4_final}). This is actually the subject of the present lecture: to understand how it is possible to reconstruct high angular resolution images of a distant celestial source using modern optical/IR interferometers such as VLTI, CHARA, etc. In fact, the image of a distant star that one would see in the focal plane of a Fizeau-type interferometer is no longer just an Airy disk due to each single telescope aperture but a brighter Airy disk superimposed with a series of interference fringes, alternately bright and dark, perpendicularly oriented with respect to the line joining the two telescopes and with an inter-fringe angular separation equal to $\lambda / B$, where $B$ is the baseline of the interferometer (see Fig.~\ref{Fig_5_final}). This naturally leads to the hope that it will be possible to retrieve along the direction of the baseline having a length $B$ an angular resolution that is equivalent to that of a single dish telescope having a diameter $B$. As a summary, figure~\ref{Fig_6_final} illustrates the improvement expected in angular resolution while observing an extended celestial source with telescopes of increasing size ($D_2 > D_1$) and with an interferometer composed of two telescopes separated by a distance $B > D$.


\begin{figure}[h]
\sidecaption
\centering
\includegraphics[width=9cm]{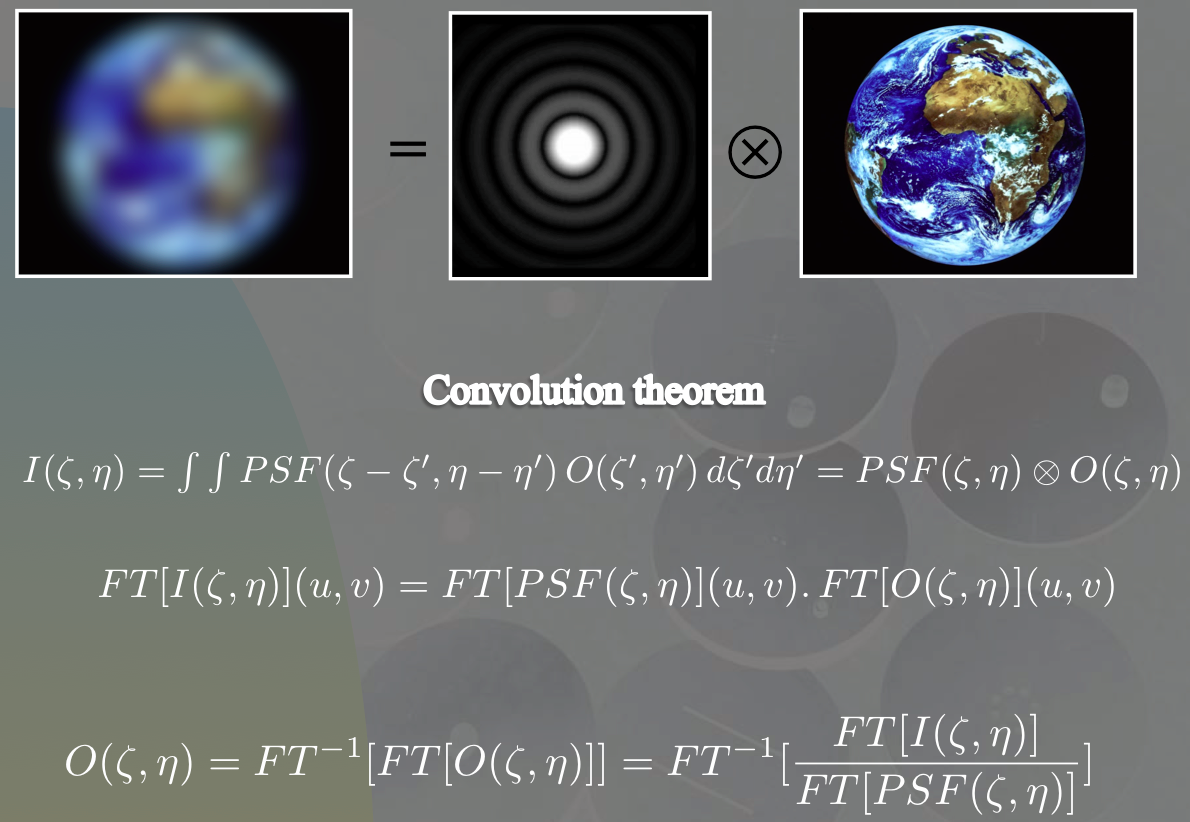}
\caption{The image I($\zeta,\eta$) we observe in the focal plane of an instrument (cf. single dish telescope) from a distant extended source as a function of its angular coordinates $\zeta,\eta$ is the convolution product of the real source image (cf. the extended Earth-like planet, O($\zeta,\eta$)) by the point spread function PSF($\zeta,\eta$) of the telescope. }
\label{Fig_7_final}       
\end{figure}

In mathematical terms, the convolution theorem states that the image I($\zeta,\eta$) we observe in the focal plane of an instrument (single dish telescope or interferometer) from a distant extended source as a function of its angular coordinates $\zeta,\eta$ is the convolution product of the real source image (cf. the extended Earth-like planet), O($\zeta,\eta$) by the point spread function PSF($\zeta,\eta$) of the telescope (i.e. the Airy disk, see Fig.~\ref{Fig_7_final}) or of the interferometer (i.e. the Airy disk crossed by the interference fringes). 

While taking the Fourier transform (FT) of the first expression given in Fig.~\ref{Fig_7_final}, we find that $FT[I(\zeta,\eta)](u,v)$ is simply equal to the natural product of $FT[PSF(\zeta,\eta)](u,v)$ and $FT[O(\zeta,\eta)](u,v)$ where $u, v$ represent the angular space frequencies defined as $u = B_u / \lambda$ and $v = B_v / \lambda$, respectively, where $B_u$ and $B_v$ correspond to the projected baselines of the interferometer along the directions parallel to the angles $\zeta,\eta$. One can then expect that by just taking the inverse Fourier transform $FT^{-1}$ of $FT[O(\zeta,\eta)](u,v)$, it will become possible to retrieve high angular resolution information about the extended source with an angular resolution equivalent to $1/u = \lambda / B_u$ and $1/v = \lambda / B_v$, respectively:
\begin{equation}
\label{eq:1}
O(\zeta,\eta) = FT^{-1}[FT[O(\zeta,\eta)](u,v)](\zeta,\eta) = FT^{-1}[\frac{FT[I(\zeta,\eta)](u,v)}{FT[PSF(\zeta,\eta)](u,v)}](\zeta,\eta).
\end{equation}
The quantity $FT[I(\zeta,\eta)](u,v)$ can be directly derived from the observation of the extended source with the optical/IR interferometer while the other quantity $FT[PSF(\zeta,\eta)](u,v)$ can be obtained from the observation of a point-like (unresolved) star. During this lecture, we shall see that the Wiener-Khinchin theorem states that the latter quantity is also merely given by the auto-correlation function of the distribution of the complex amplitude of the radiation field in the pupil plane of the observing instrument being used (single dish telescope or interferometer). The goal of the present lecture is to establish relations such as Eq.~(\ref{eq:1}).  

\section{Some reminders}

With a few exceptions (cf. the Moon, the Sun, the Andromeda Galaxy, etc.), all the celestial objects that we see in the sky appear to us, with the naked eye, as point-like objects. Apart from their apparent motion with respect to the fixed stars on the celestial sphere, we are not even able to distinguish between the images of Jupiter, Saturn or even Venus from those of ordinary stars.
We describe in this course an observation method based on the principle of a Fizeau-type interferometer, which allows with just some basic cooking equipment to resolve angularly a planet such as Venus, when it is at its maximum apparent brightness ($V \sim -4.4$).

\begin{figure}[h]
\sidecaption
\centering
\includegraphics[width=9cm]{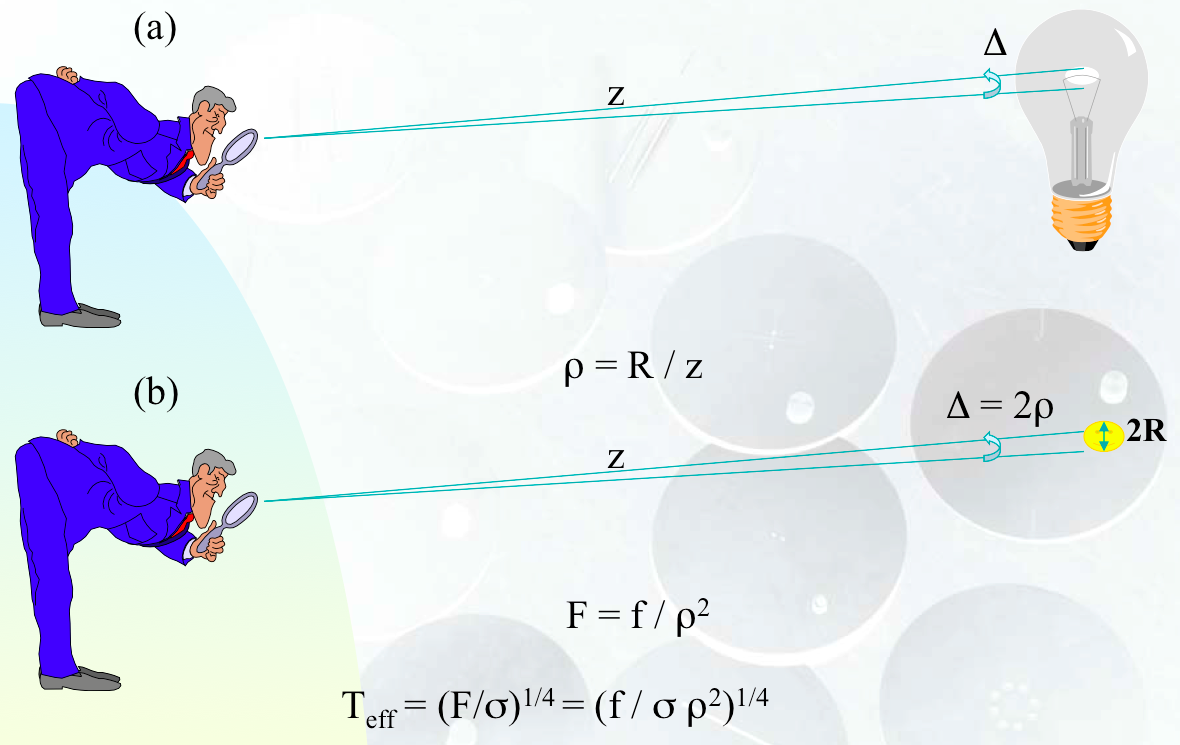}
\caption{Resolving the angular diameter of a star (b) is alike trying to estimate the angular size of the filament of a light bulb (a). }
\label{Fig_8_final}       
\end{figure}

If we assimilate for a moment the disc of a star, or even that of Venus, to the filament of a light bulb, the object of the present lecture can still be formulated as follows: given a common electric light bulb inside which is a filament, having a certain thickness $T$ (measured perpendicularly to the line-of-sight) and which is incandescent (cf. a star), how to measure the thickness $T$ of this filament (diameter of the star) not only without breaking the bulb but also assuming that it is so far away from us, at a distance $z$, that it is not possible for us to angularly resolve the filament with the naked eye (see Fig.~\ref{Fig_8_final}a)?

Let us now recall that knowledge of the angular radius ($\rho=R / z$) of a star located at a distance $z$ ($z >> R$) and having a linear radius $R$ allows the direct determination of its flux $F$ at the stellar surface from the flux $f$ observed on Earth (as a reminder $F = f / \rho^2$, see Fig.~\ref{Fig_8_final}b). If we can measure the absolute distance $z$ of the star, we can also determine its linear radius $R$ from its angular radius $\rho \thinspace (R = \rho z)$. Moreover, knowledge of the intrinsic flux $F$ of the star allows an immediate determination of its effective temperature $T_{eff}$, thanks to the application of the Stefan-Boltzmann law ($F = \sigma T_{eff}^4$). It then results that $T_{eff} = (f / \sigma \rho^2)^{1/4}$. 
The measurements of the angular radius and of the flux of a star measured on Earth thus lead to the determination of the effective temperature $T_{eff} $ of that star. As a reminder, this temperature is directly involved in the construction of stellar atmosphere models and stellar evolution.
   We will also show that Fizeau-type stellar interferometry literally allows direct imaging with very high angular resolution of distant bodies by the method of aperture synthesis.
   Let us now proceed with a few theoretical reminders about the description of a field of electromagnetic light radiation.

\subsection{Complex representation of an electromagnetic wave}

Let us first remind that a beam of light radiation can be assimilated to the propagation of a multitude of electromagnetic waves at the speed of 299,792 $km \thinspace s^{-1}$ in the vacuum.
   If, for the sake of simplicity, we assume that we deal with a plane monochromatic wave, linearly polarized, propagating along the direction of abscissa $z$, the electric field $E$ at any point in space and at time $t$, can be represented by a sinusoidal type function taking for example the shape

\begin{equation}
\label{eq:2}
E = a \hspace{0.1 cm} cos (2\pi(\nu t - z / \lambda))
\end{equation}
where


\begin{equation}
\label{eq:3}
\lambda = c \hspace{0.1 cm} T = c / \nu 
\end{equation}

\begin{figure}[h]
\sidecaption
\centering
\includegraphics[width=6cm]{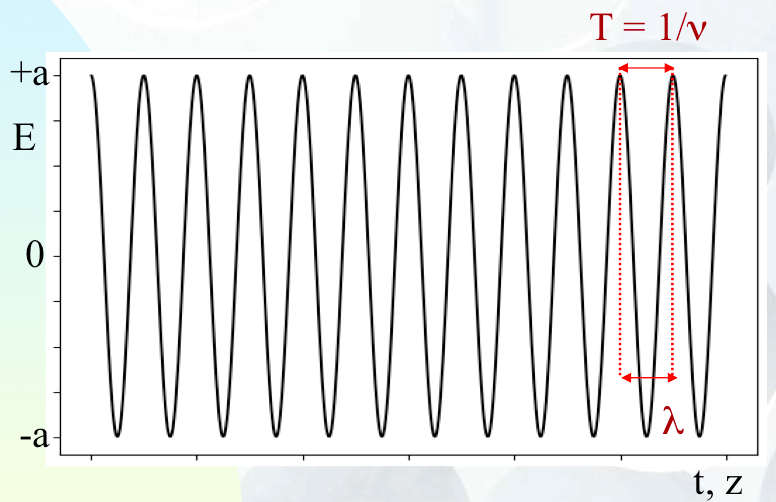}
\caption{Representation of an electromagnetic wave.} 
\label{Fig_9_final}    
\end{figure}

$c$, $\lambda$, $\nu$, $T$ and $a$ representing the speed of light, the wavelength, the frequency, the period and the amplitude of the electromagnetic vibrations, respectively (see Figure \ref{Fig_9_final}). 

 We know how convenient it is to rewrite the previous equation in complex notation:
\begin{equation}
\label{eq:4}
E = Re \{a  \thinspace \exp [i2\pi(\nu t - z / \lambda)]\}
\end{equation}

where $Re$ represents the real part of the expression between the two curly braces. This complex representation of an electromagnetic wave has the great advantage that the exponential function can now be expressed as the product of two functions depending separately on the spatial and temporal coordinates 

\begin{equation}
\label{eq:5}
E = a  \thinspace \exp (-i\phi) \thinspace \exp(i2\pi\nu t)
\end{equation}
where
\begin{equation}
\label{eq:6}
\phi = 2\pi z / \lambda.
\end{equation}

If we suppose that all the operations that we carry out on the electric field $E$ are linear, it is of course very convenient to use in our calculations its complex representation (see Eq.(\ref{eq:5})) and to take at the end the real part of the result obtained.

   We can then rewrite the previous equation as follows:
\begin{equation}
\label{eq:8}
E = A  \thinspace \exp(i2\pi\nu t),
\end{equation}
where
\begin{equation}
\label{eq:9}
A = a  \thinspace \exp (-i\phi)
\end{equation}
with A representing the complex amplitude of the vibration.

Because of the extremely high frequencies of electromagnetic waves corresponding to visible radiations ($\nu \sim 6 \thinspace10^{14} Hz$ for $\lambda = 5000 \thinspace \AA$), we recall that it is not normally possible to make direct observations of the electric field $E$ (the situation is different in the radio domain). The only measurable quantity is the intensity $I$, which is the time average of the amount of energy passing through a unit surface element, per unit of time and solid angle, placed perpendicularly to the direction of propagation of the light beam.

The intensity $I$ is therefore proportional to the temporal average of the square of the electric field:
\begin{equation}
\label{eq:10}
\left\langle E^2 \right\rangle = lim_{T \rightarrow \infty} \frac{1}{2T} \int_{-T}^{+T} E^2 dt,
\end{equation}

which is reduced to (e.g. replace in the previous relation $E$ by Eq.(\ref{eq:2}))
\begin{equation}
\label{eq:11}
\left\langle E^2 \right\rangle = \frac{a^2}{2},
\end{equation}

where $a$ is the real amplitude of the electric field.

By convention, the intensity of the radiation is defined by the following relation:
\begin{equation}
\label{eq:12}
I = A \hspace{0.1cm}A^{\ast} = |A|^2 = a^2.
\end{equation}

\subsection{Principle of Huygens-Fresnel }

\begin{figure}[h]
\sidecaption
\centering
\includegraphics[width=9cm]{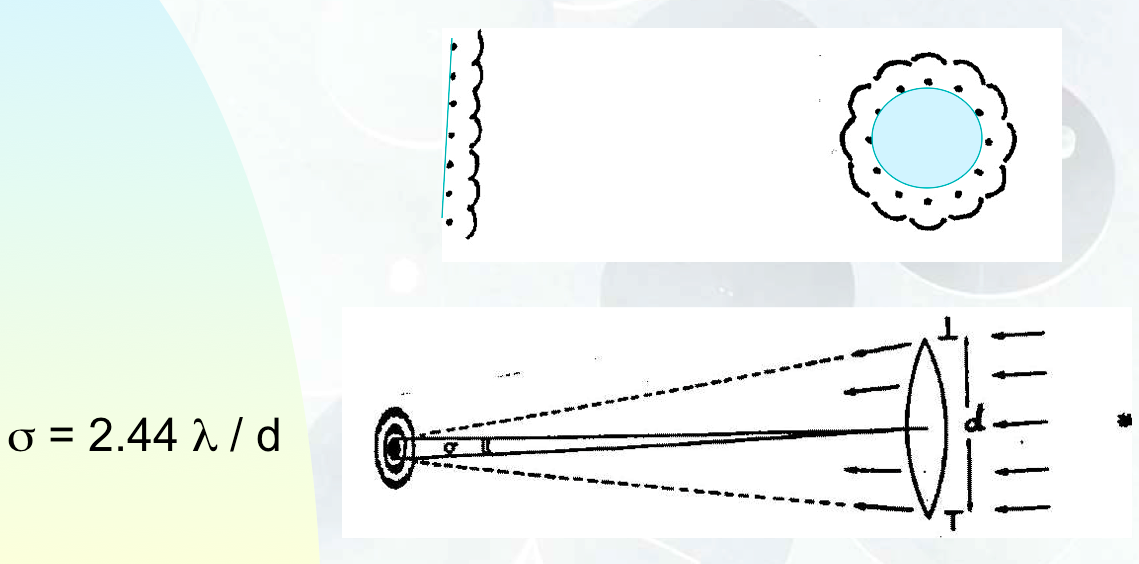}
\caption{Illustration of the Huygens-Fresnel principle during the propagation of a plane or circular wavefront and diffraction of light which encounters a converging lens. } 
\label{Fig_10_final}   
\end{figure}

We recall that, according to Huygens, each point of a wavefront can be considered as being the centre of a secondary wave leading to the formation of spherical wavelets, and that the main wavefront, at any subsequent moment can be considered as the envelope of all these wavelets (see Fig.~\ref{Fig_10_final}). 

Using this model, Fresnel was the first to account for the observed effects of light diffraction, assuming that secondary wavelets interfere with each other. This combination of the Huygens construction method and the Fresnel interference principle is called the Huygens-Fresnel principle. This is the basis of the concept of the Fourier transform. Let us remind a direct application of this principle when studying the formation of the image of a distant object at the focus of a telescope having a linear diameter $d$. Following the diffraction of the waves at the passage of the opening of the telescope (as if the waves were trying to spread and bypass the obstacles), we observe a phenomenon of redistribution of the energy of the light wave: the image of a point-like source produced by a converging circular objective (lens or mirror) is not a point but spreads in a diffraction pattern called the "Airy disk" (see Fig.~\ref{Fig_10_final}). The angular diameter of the central spot is (in radian):
\begin{equation}
\label{eq:13}
\sigma = 2.44 \thinspace \lambda / d
\end{equation}
where $\lambda$ is the wavelength of light and $d$ is the linear diameter of the aperture.

\begin{figure}[h]
\sidecaption
\centering
\includegraphics[width=7cm]{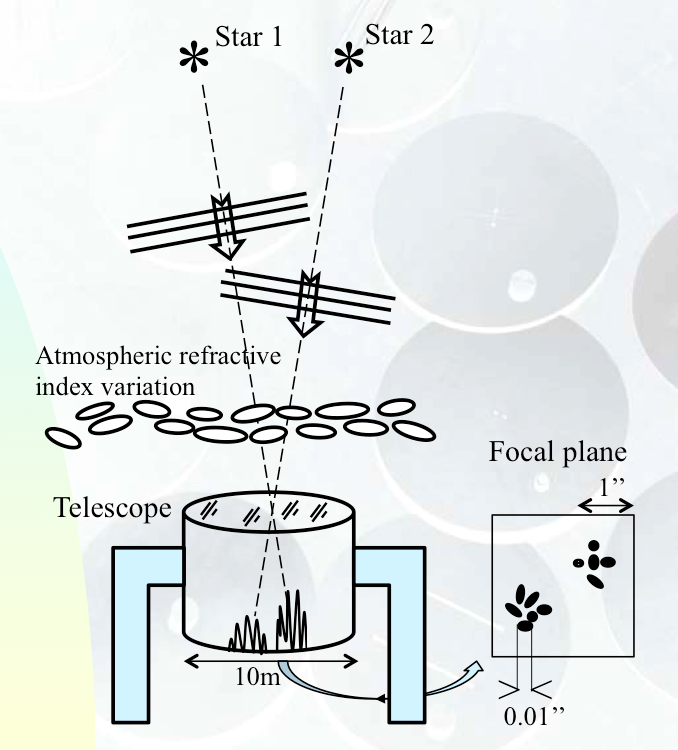}
\caption{Atmospheric agitation above the objective of a large telescope causing the seeing effects seen in its focal plane. } 
\label{Fig_11_final}   
\end{figure}

We can resolve an extended source by direct imaging, if and only if, its angular diameter $\Delta$ (= $2\rho$) is somewhat larger than $\sigma$. For example, our pupil whose approximate diameter varies between 1 and 5 mm, allows us to angularly resolve nearby objects separated by more than 138'' and 28'', respectively. In the visible range, a telescope, with a diameter of 14 cm, will allow us to resolve objects with an angular dimension larger than 1'', and for diameters larger than 14 cm, their collecting area will naturally be enhanced but their angular resolution will remain limited to (more or less) 1'' because of the atmospheric agitation (see Fig.~\ref{Fig_11_final}). In fact, under the influence of temperature and pressure gradients, a regime of eddies establish itself in the Earth atmosphere which, at low altitude ($\sim$ 10 km), have dimensions of the order of 20 cm (sometimes only a few cm, sometimes 30 or 40 cm) and evolution periods of the order of a few milliseconds. Optically, these changes manifest themselves by an inhomogeneity in the refractive index distribution. The amplitude and the relative phase shift of the electromagnetic field in the pupil plane thus get disturbed in a random manner. 

It follows that if we observe the Moon, Jupiter, etc. either with the largest telescope in the world with a diameter of 10m ($\sigma$ = 0.014'') or with an amateur telescope with a diameter of 14 cm ($\sigma$ = 1''), we will see the same details under good seeing conditions. The brightness will of course be larger with the 10m telescope ... but probably too bright for the eye not to be blinded by the image of the Moon or Jupiter.

The diffusion indicator of the atmosphere is defined as being the average inclination perturbation of the wave surfaces. This tilt disturbance reaches values that vary between 1'' and 10'', depending on the site and the moment.
   This phenomenon is detected differently according to the dimensions of the instrument used. The eye, which has an angular resolution close to the minute of arc, will be sensitive only to variations of amplitude: we then see stars flickering. An instrument of 10 to 20 cm in diameter will detect tilt variations and the focal image will oscillate around an average position. For larger instruments, a large number of eddies will, at the same time, be involved in the formation of the focal image. This will therefore have the dimensions of the diffusion indicator of the atmosphere. The spatial coherence of the entrance pupil will allow, for a point-like source, the realization of interference phenomena between the radiations passing through different points of the pupil. A statistical study makes it possible to show that the resulting focal image, delimited by the diffusion indicator, consists of a set of granules (called 'speckles') which have the size of the Airy disk of the instrument (see Fig.~\ref{Fig_11_final}). These granules swarm in the diffusion spot at the rhythm of the change of the atmospheric eddies. The stability of the focal image is therefore also of the order of the millisecond. The technique of speckle interferometry, developed by the French astronomer Antoine Labeyrie, allows to re-construct the images of the stars observed with the angular resolution given by the true diameter of the telescope.

\section{Brief history about the measurements of stellar diameters}
In the past, there have been numerous attempts to measure angular diameters of stars, and we will first recall three of these approaches that clearly show the difficulties encountered.

\subsection{Galileo}

\begin{figure}[h]
\sidecaption
\centering
\includegraphics[width=8cm]{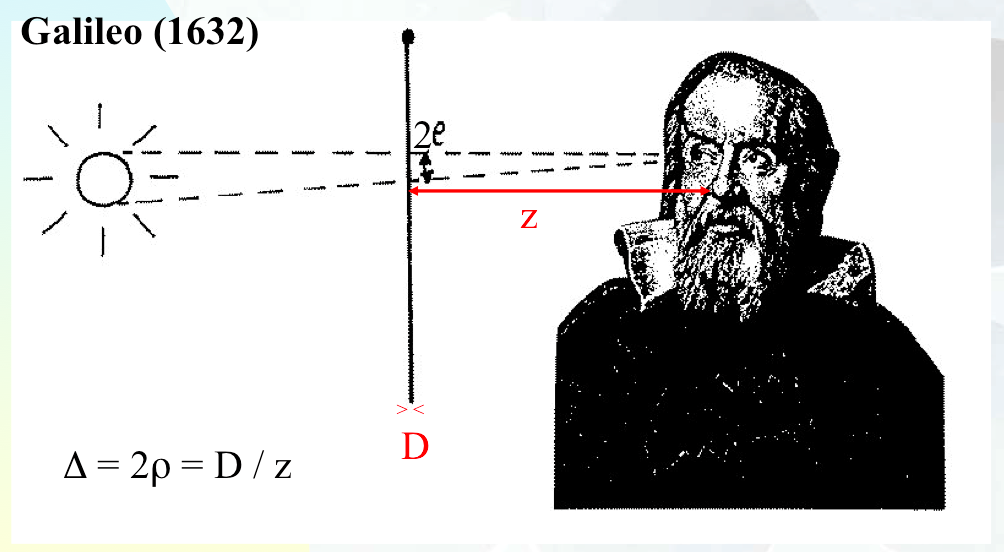}
\caption{Experimental measurement by Galileo of the angular diameter of a star (see text).} 
\label{Fig_12_final}   
\end{figure}

A first experimental attempt to measure the angular diameter of stars was made by Galileo (1632). He proceeded as follows: placing himself behind a rigid wire (whose thickness $D$ was known, see Fig.~\ref{Fig_12_final}) suspended vertically, he determined the distance $z$ to which he had to move in so that the image of the star Vega ($\alpha$ Lyrae) of magnitude zero got completely obscured by the wire. Galileo deduced that the angular diameter of Vega, equal to that of the wire, was about 5'', which was in itself a rather revolutionary result, since the value adopted at that epoch for the angular diameter of the stars was close to 2'. As we saw earlier, the value of 2' is certainly the result of the low angular resolution of our eye, while the 5'' angular diameter measured by Galileo was the result of the effects of the atmospheric agitation (seeing effects) at the time of his observations.

\subsection{Newton}
A theoretical estimate of the angular dimension of a star of magnitude zero was performed by Newton. His approach was as follows: if we suppose that the Sun is a star similar to the stars situated on the celestial sphere and if we place our star of the day at a distance $z$ such that its apparent brightness $V_\odot$  becomes comparable to that of a star of magnitude equal to zero, then its angular diameter $\Delta$ should be of the order of $2 \thinspace 10^{-3}$'' (with the current value of the visual apparent magnitude of the Sun, $V_\odot$  = -26.7, we find $\sim 8 \thinspace 10{-3}$''). It should be noted that the value currently established for the star Vega with modern interferometers is $3 \thinspace 10^{-3}$''. The formula to be used to establish this result can be obtained as follows: the angular diameter of the Sun $\Delta$ placed at the distance of Vega ($V$ = 0) is given by the product of the apparent angular diameter of the Sun $\Delta_\odot$ times the factor $10^{V_\odot / 5}$. As a reminder, the apparent diameter of the Sun is about 30'.

\subsection{Fizeau-type interferometry }

The third experimental attempt of measuring stellar diameters, based on Fizeau-type interferometry, is in fact the work of prominent scientists such as Young, Fizeau, Stephan, Michelson and Pease. These last two having measured the first angular diameter of a star in 1920. Although other methods of interferometric measurements of stellar angular diameters appeared later (cf. the interferometry in intensity of Brown and Twiss in 1957, speckle interferometry by Antoine Labeyrie in 1970, etc.), we will only describe in detail the Fizeau-type interferometry, which is still the most powerful technique used and the most promising measurement of angular diameters of stars and imagery at very high angular resolution of distant bodies by the aperture synthesis method.

\begin{figure}[h]
\sidecaption
\centering
\includegraphics[width=9cm]{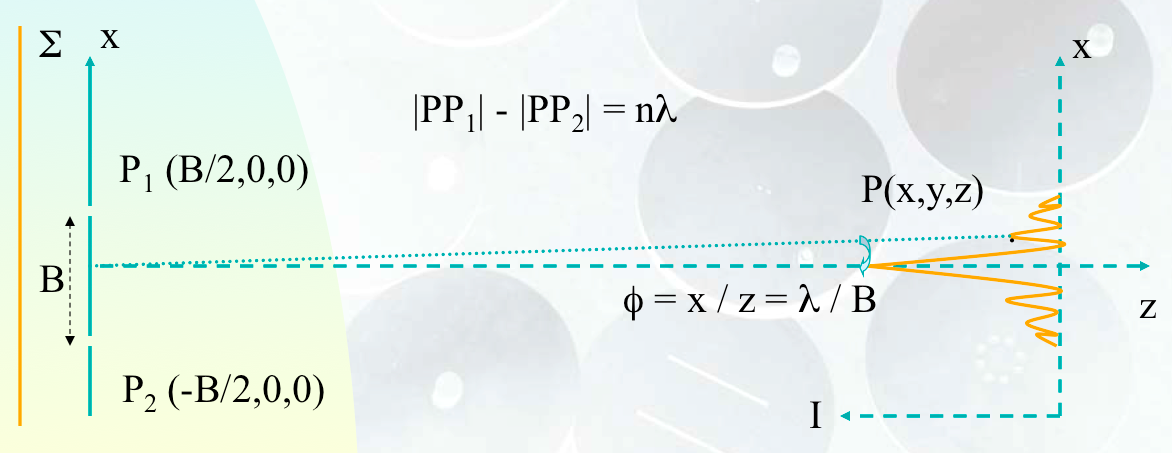}
\caption{The  double hole experiment of Young (see text).} 
\label{Fig_13_final}   
\end{figure}

   Let us first remind the results obtained in the Young double hole experiment (1803, see Fig.~\ref{Fig_13_final}). 

A monochromatic plane wave coming from a distant point-like source is falling on a screen drilled with two holes ($P_1$ and $P_2$) separated along the $x$ axis by a baseline $B$. In accordance with the Huygens-Fresnel principle, the two holes will emit spherical waves that will interfere constructively whenever the difference in their propagation lengths is a multiple of $\lambda$ (see Eq.~(\ref{eq:14}) below), and destructively if it corresponds to an odd number of half wavelengths.

   The locus of points $P(x, y, z)$ with cartesian coordinates $x$, $y$, $z$ (see Fig.~\ref{Fig_13_final}) where there will be a constructive interference is thus given by 

\begin{equation}
\label{eq:14}
|P_1P| - |P_2P| = n \lambda
\end{equation}

with $n = 0, \pm1, \pm2, $ etc.

\begin{figure}[h]
\sidecaption
\centering
\includegraphics[width=10cm]{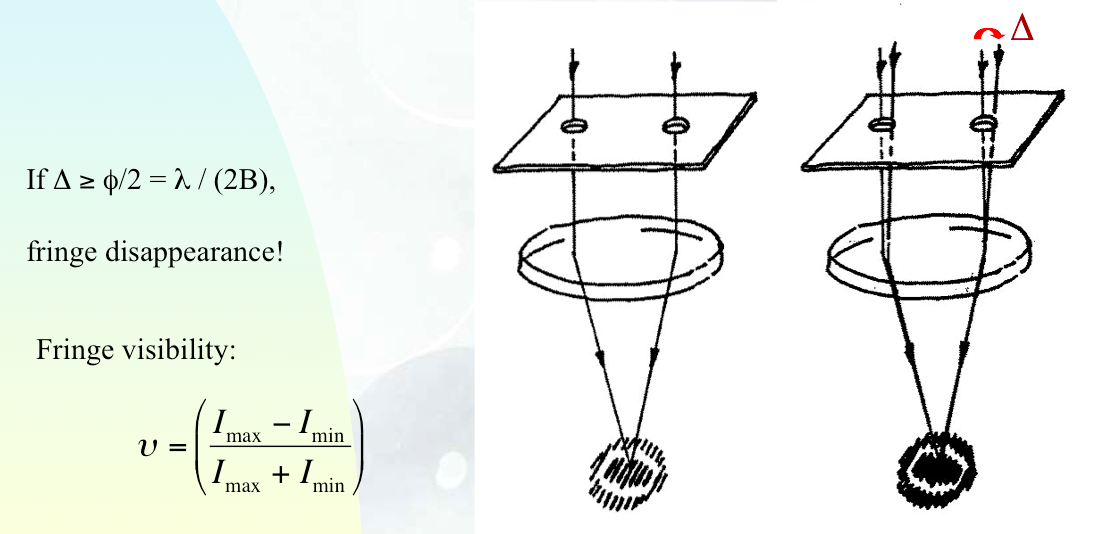}
\caption{Fizeau experiment: for the case of a single star (left drawing) and for the case of a double star with an angular separation $\Delta$  (right drawing, see text).} 
\label{Fig_14_final}   
\end{figure}

   Let the points $P_i(x_i, y_i, 0)$ in the screen plane and $P(x, y, z)$ in the observer plane be such that $|x_i|, |y_i|, |x|, |y| << |z|$. We then find that
\begin{equation}
\label{eq:15}
|P_iP| = \sqrt{(x - x_i)^2 + (y - y_i)^2 + z^2}
\end{equation}
which can be simplified at first order (given the above conditions) as follows:

\begin{equation}
\label{eq:16}
|P_iP| = z \{1 + \frac{(x - x_i)^2 + (y - y_i)^2}{2z^2}\}.
\end{equation}
Considering the two points $P_1$ and $P_2$ in the Young's screen, Eq.~(\ref{eq:14}) reduces to
\begin{equation}
\label{eq:17}
z \{1 + \frac{(x + B / 2)^2 + y^2)}{2z^2}\} - z \{1 + \frac{(x - B / 2)^2 + y^2}{2z^2}\} = n \lambda
\end{equation}
and finally
\begin{equation}
\label{eq:18}
\frac{x B}{z} = n \lambda
\end{equation}
or else
\begin{equation}
\label{eq:19}
\Phi = \frac{x}{z} = n \frac{\lambda}{B}.
\end{equation}
Since the angular separation $\Phi$ between two successive maxima (or minima) does not depend on the coordinate $y$, there results a pattern of bright and dark fringes, oriented perpendicularly with respect to the line joining the two holes, and with an inter-fringe angular separation $\Phi = \lambda / B$. In case the two holes are not infinitely small, the observed interference pattern will naturally overlap the combined Airy disks produced by each single hole. For $\lambda = 5500 \thinspace \AA $ and $B = 1$ mm, we find that $\Phi = 113$'', just at the limit of our eye visual resolution.
   In 1868, the French optician Hyppolite Fizeau realized that in the Young's hole experiment presented above, the contrast of the interference fringes decreased as the diameter of the light source widened. Similarly, it decreased when the distance $B$ between the two holes was extended. Was there a simple relation between the angular diameter $\Delta$ of the source and the spacing $B$ between the two holes corresponding to the disappearance of the fringes? Before establishing such a rigorous relationship, let us try to understand this observation intuitively on the basis of simple geometrical considerations (see Fig.~\ref{Fig_14_final}).
   Indeed, if instead of considering the diffraction pattern given by Young's holes for a single point-like source, we consider a composite source made of two incoherent point-like sources separated by an angle $\Delta$, that is to say between which there is no interference between their light, it will result in the plane of the observer a superposition of two systems of Young fringes, separated by an angle $\Delta$. If $\Delta \sim \Phi / 2$, there will result a total scrambling of the fringes. The bright fringes of one source will overlap the dark fringes of the second one and their contrast will totally vanish.

From Fig.~\ref{Fig_14_final}, it is clear that the visibility of the fringes will significantly decrease whenever the following condition takes place  
\begin{equation}
\label{eq:20}
\Delta > \frac{\Phi}{2} = \frac{\lambda}{2B}.
\end{equation}

   A quantity that objectively measures the contrast of the fringes is called the visibility. It is defined by the following expression:
\begin{equation}
\label{eq:21}
v=\left(\dfrac{I_{\max}-I_{\min}}{I_{\max}+I_{\min}}\right).
\end{equation}
Whenever a star is not resolved, we have $I_{min} = 0$, and thus the visibility $v =1$. If the star is being resolved, $I_{max} = I_{min}$  and thus the resulting visibility $v = 0$. 

   Fizeau proposed in 1868 to apply this method to stellar sources. He found sufficient to place a screen drilled with two elongated apertures at the entrance of a telescope pointed towards a star and to look in the focal plane by means of a very powerful eyepiece the Airy disk crossed by the Young's fringes and to increase the distance between the two apertures until the visibility of the fringes vanishes.

\begin{figure}[h]
\sidecaption
\centering
\includegraphics[width=5cm]{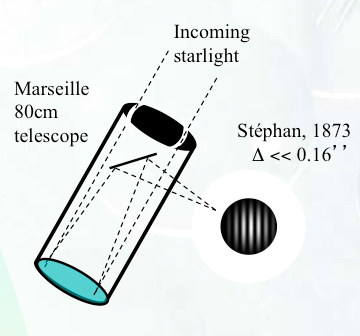}
\caption{Diagram illustrating the way Fizeau and Stephan proceeded in order to measure the angular diameters of stars with the interferometric technique. } 
\label{Fig_15_final}   
\end{figure}

\begin{figure}[h]
\sidecaption
\centering
\includegraphics[width=5cm]{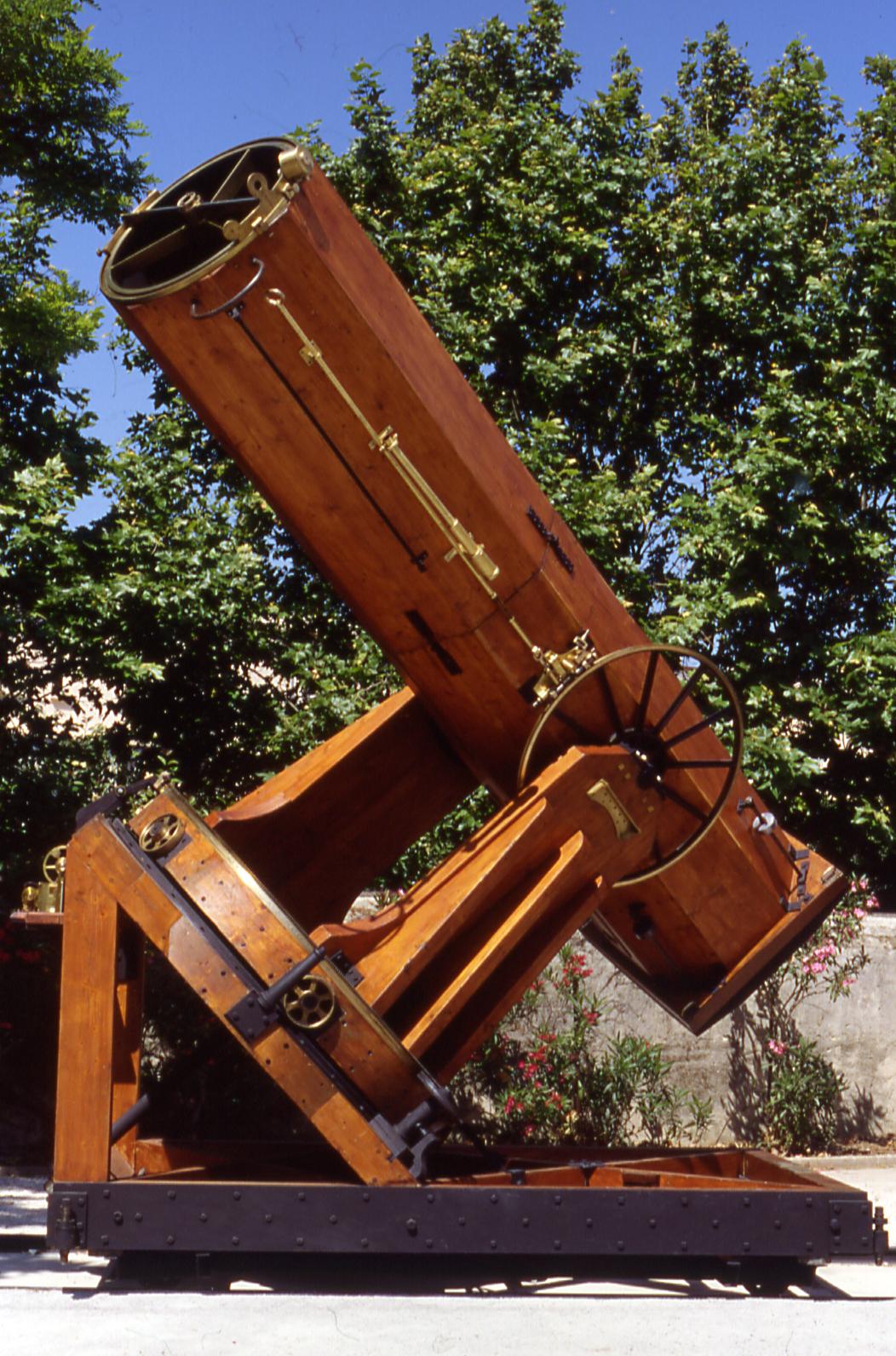}
\caption{The 80cm Marseille telescope used by Fizeau and Stephan. \copyright \thinspace Michel Marcelin.} 
\label{Fig_16_final}   
\end{figure}

   This experiment is attempted in 1873 by Stephan with the 80cm telescope of the Marseille Observatory. All the bright stars visible in the sky are observed. The two openings at the entrance were actually in the form of crescents but one may demonstrate that the contrast of the fringes is independent of the shape of the two openings if they are identical. The result was disappointing: with a maximum base separation of 65cm between the openings, no attenuation of the contrast of the fringes was observed for any star. This proved that no star could be resolved using that instrument. Stephan concluded that the angular diameter of the stars is much smaller than 0.16'' (see Fig.~\ref{Fig_15_final}). From these observations, it is of course possible to set a lower limit on the effective temperature $T_{eff}$ of all those stars (see Section 2). Figure \ref{Fig_16_final} illustrates the 80 cm Marseille telescope used by Stephan and Fizeau. 

\subsection{Home experiments: visualization of the Airy disk and the Young interference fringes}

\begin{figure}[h]
\sidecaption
\centering
\includegraphics[width=7cm]{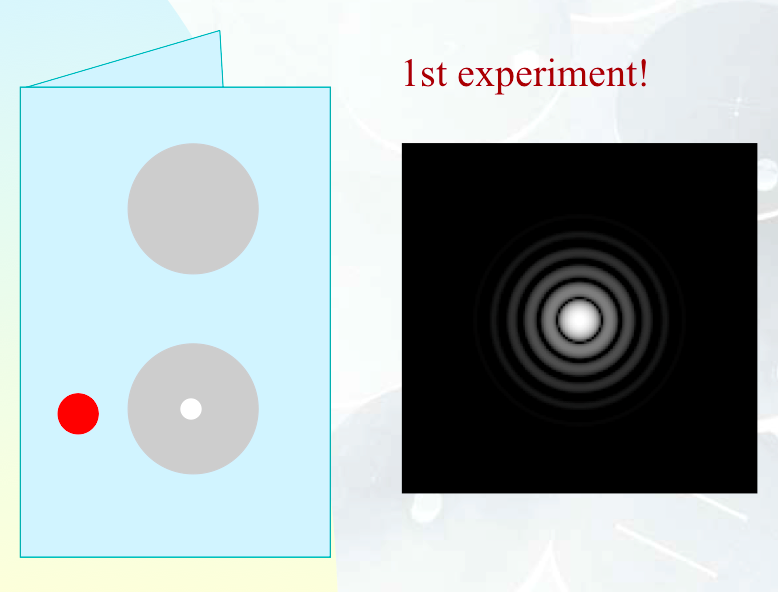}
\caption{The one hole screen experiment: the small circular hole drilled in the aluminium paper is visible inside the lower bigger hole perforated in the cartoon screen. When looking through this hole at a distant light bulb, you perceive a nice Airy disk (cf. right image). } 
\label{Fig_17_final}   
\end{figure}

We propose hereafter two simple experiments that can be carried out at home in order to visualize with our own eyes the Airy disk and the interference fringe patterns. 
   To do this, take a rectangular piece of cartoon ($\sim$ 5 cm x 10 cm) and fold it in the middle (see Fig.~\ref{Fig_17_final}). Perforate through the two sides of this piece of cartoon two well separated circular holes having an approximate diameter of 1/2 cm.  
Cut a thin sheet of aluminium paper into two small squares (cf. 1cm x 1cm). 
In one of the two squares, drill with a thin metallic pin a $\sim $ 0.5 mm (or smaller) diameter circular hole near its centre. Glue inside the folded cartoon the first aluminium piece in such a way that the very small hole is centered with respect to one of the big holes (cf. the lower one) of the cartoon.

\begin{figure}[h]
\sidecaption
\centering
\includegraphics[width=7cm]{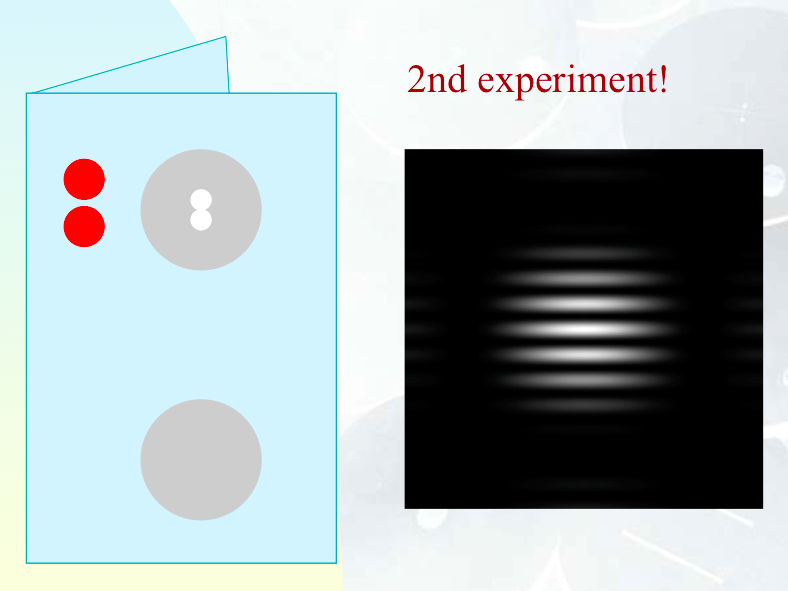}
\caption{The two hole screen experiment: the two small circular holes drilled in the aluminium paper are visible inside the upper bigger hole perforated in the cartoon screen. When looking through these two holes at a distant light bulb, you perceive a nice Airy disk superimposed by a pattern of bright and dark fringes (cf. right image).} 
\label{Fig_18_final}   
\end{figure}

\begin{figure}[h]
\sidecaption
\centering
\includegraphics[width=9cm]{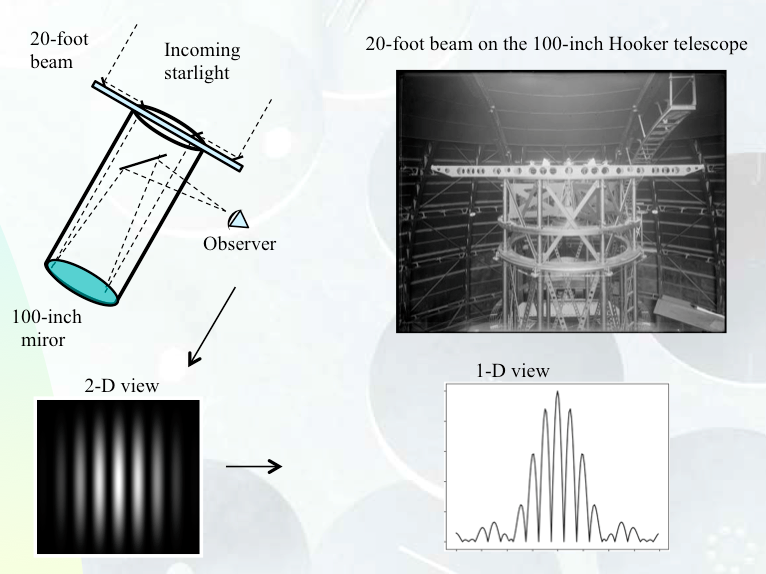}
\caption{The stellar interferometer of Michelson and Pease set on top of the 2.5m Mount Wilson telescope. \copyright \thinspace The Observatories of the Carnegie Institution.} 
\label{Fig_19_final}   
\end{figure}

Take the second square of aluminium and drill with the thin metallic pin two 0.5 mm (or smaller) circular holes near its centre separated by about 0.5-1 mm (see Fig.~\ref{Fig_18_final}).  
Glue now near the second circular hole (cf. the upper one) inside the folded cartoon this second aluminium square. After, you should glue the two sides of the cartoon in such a way that you can hold it with ease in one hand. Of course you can do this on a single cartoon or on two separate ones.
Place at a distance of about 10 m a small light bulb (cf. the light of a cell phone) and look at it through the single hole drilled in the aluminium. You should see a nice Airy disk which angular diameter is $2.44 \thinspace \lambda / D$, with $D$ being about 0.5 mm and $\lambda$ the wavelength of the ambient light ($\sim  5000 \thinspace \AA$). Now looking at the distant light source through the double hole drilled in the other square of aluminium, you should see an Airy disk superimposed by a series of white and dark fringes, oriented perpendicularly with respect to the line joining the two holes, with an angular inter-fringe separation $\lambda / D$, where $D \sim$ 0.5-1 mm. If you rotate azimuthally the screen, you will observe that the fringes also rotate since they constantly remain perpendicular to the line joining the two small holes of the milli-interferometer. While getting closer to the small light bulb, you will notice that the visibility of the fringes decreases. Let $Dist$ be the distance around which the latter totally vanishes. The product of $Dist$  by  $\lambda / D$ corresponds to the linear diameter of the light bulb. Instead of changing the distance between the light bulb and the milli-interferometer, one could change the separation between the two holes and determine the separation for which the interference fringes disappear.
Adopting the same approach as Stephan, Abraham Michelson used in 1890 the Lick 30cm telescope to resolve the four Galilean satellites of Jupiter. Their angular sizes were of the order of 0.8'' - 1.4'' while the resolving power provided by the largest baseline he used was about 0.5''. An excellent agreement was found for the angular diameters of the satellites with the classical measurements made at the same time. 
   To resolve the biggest stars, much longer baselines are needed. Michelson and Pease built a 7m metal beam carrying four 15cm flat mirrors that they installed at the top of the Mount Wilson telescope, having a diameter of 2.5m (see Fig.~\ref{Fig_19_final}). 

The two mobile exterior mirrors formed the basis of the interferometer and the two fixed interior mirrors returned the star's light into the telescope. With a maximum baseline of 7m, the smallest measurable angular diameter was 0.02''. Use of this first stellar interferometer was very delicate because the visualization of the Young fringes was only possible if the two optical paths from the star passing through the two exterior mirrors and reaching the observer eyes were equal to an accuracy of about 2 microns (see discussion below).
Michelson and Pease finally obtained the first measurement of a stellar diameter during the winter of 1920, that of Betelgeuse ($\alpha$ Orionis), a red supergiant. They found an angular diameter of 0.047'', that is a linear diameter 400 times larger than that of the Sun, given the distance of Betelgeuse (650 light years). Five more bright stars were also resolved.
   Anderson used the same observing technique with the 2.5m telescope at Mount Wilson to resolve very tight spectroscopic binaries (cf. Capella). Michelson and Pease did not stop there: they undertook the construction of a 15m optical beam based on the same principle, and began to use it in 1929. Unfortunately, the mechanical vibrations and deformations were such that this instrument was too delicate. It was abandoned in 1930, without having reached its limiting angular resolution of 0.01''.
   It was not until 1956 that optical stellar interferometry was reborn and again, according to a principle different from that of Fizeau. Fizeau-type interferometry had indeed acquired a reputation of great operational difficulty. The intensity interferometry by the two radar manufacturers Hanbury Brown and Twiss (Australia) was then set up, based on an entirely new approach: the measurement of the space correlation of the stellar intensity fluctuations. Their interferometer made it possible to measure the diameter of 32 blue stars with a very high precision ($<$ 0.0005'') and to detect a few very tight binaries. But it is in the field of radio astronomy that the development of interferometry with independent telescopes became the most spectacular in the 1950s.
   At optical wavelengths, we had to wait until 1975 when Prof. Antoine Labeyrie and his close collaborators succeeded in combining the light from two independent telescopes. A boost then took place in the successful development of optical/IR interferometry. 

\section{Light coherence}

When we previously established the relationship between the angular diameter $\Delta$ of a source and the separation $B$ between the two apertures of the interferometer for which the interference fringes disappear, we made two approximations that do not really apply to usual conditions of observation. We first assumed that the waves falling on Young's screen were planes, that is to say coming from a very distant point-like source and also that they were purely monochromatic. In addition, the holes through which light is being scattered should have finite dimensions. 
   Therefore, we shall later take into account the finite dimensions of the apertures (see Section 6) but let us first consider the effects due to the finite dimension of the source, also considering a spectral range having a certain width and to do so, we shall make use of some elements of the theory of light coherence. This theory consists essentially in a statistical description of the properties of the radiation field in terms of the correlation between electromagnetic vibrations at different points in the field.

\subsection{Quasi-monochromatic light}

\begin{figure}[h]
\sidecaption
\centering
\includegraphics[width=8cm]{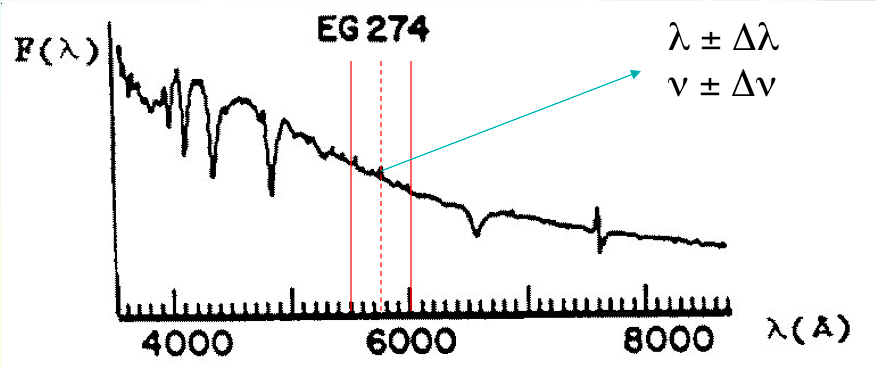}
\caption{ Stars do not emit monochromatic light. Quasi monochromatic light is assumed to be emitted at the wavelength $\lambda$ (resp. the frequency $\nu$) within the bandwidth $\pm \Delta \lambda$  (resp. $\pm \Delta \nu$). } 
\label{Fig_20_final}   
\end{figure}

The light emitted by a real source (see Fig.~\ref{Fig_20_final}) is of course not monochromatic. As in the case of a monochromatic wave, the intensity of such a radiation field at any point in space is defined by

\begin{equation}
\label{eq:22}
I = \left\langle V(t) V(t)^\ast \right\rangle.
\end{equation}

In order to determine the electric field created by such a source, emitting within a certain frequency range $\pm \Delta \nu$, we must sum up the fields due to all the individual monochromatic components such that the resulting electric field $V(z, t)$ is given by the real part of the following expression:

\begin{equation}
\label{eq:23}
V(z,t) = \int_{\nu-\Delta\nu}^{\nu+\Delta\nu} a(\nu') \thinspace \exp [i2\pi(\nu' t - z / \lambda')] d\nu'.
\end{equation}

   While a monochromatic beam of radiation corresponds to an infinitely long wave train, it can easily be shown that the superposition of multiple infinitely long wave trains, with nearly similar frequencies, results in the formation of wave groups. Indeed, the expression of the electric field established in Eq.~(\ref{eq:23})  can be reduced as follows. Insert in the integral of Eq.~(\ref{eq:23}) the following factors: $\exp(i2\pi(\nu t - z/\lambda)) \thinspace \exp(-i2\pi(\nu t - z/\lambda))$. We then find that Eq.~(\ref{eq:23}) may be rewritten as
   
\begin{equation}
\label{eq:24}
V(z,t) = A(z,t) \thinspace \exp [i2\pi(\nu t - z /\lambda)]
\end{equation}
where

\begin{equation}
\label{eq:25}
A(z,t) = \int_{\nu-\Delta\nu}^{\nu+\Delta\nu} a(\nu') \thinspace \exp \{i2\pi[(\nu'-\nu) t - z (1/\lambda'-1/\lambda)]\} d\nu'.
\end{equation}

\begin{figure}[h]
\sidecaption
\centering
\includegraphics[width=9cm]{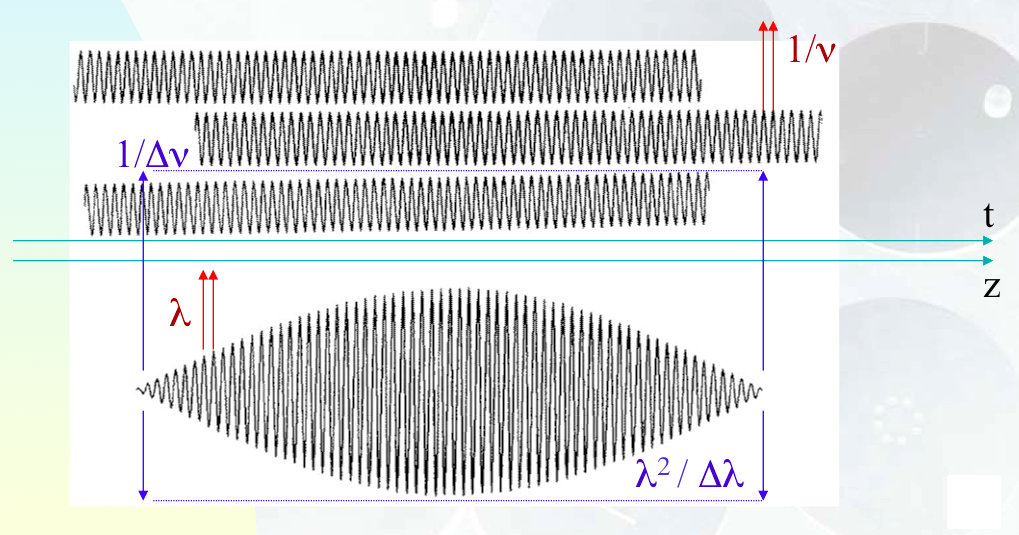}
\caption{Superposition of long wave trains having quite similar frequencies $\nu'$ in the range $\nu \pm \Delta \nu$  (resp. wavelengths $\lambda'$ in the range $\lambda \pm \Delta \lambda$) results in the propagation of long wave trains with the frequency $\nu$ (resp. wavelength $\lambda$) but which amplitude $A(z,t)$ is varying with a lower frequency $\Delta \nu$ (resp. longer wavelength $\lambda^2 / \Delta \lambda$).} 
\label{Fig_21_final}   
\end{figure}

Expression (\ref{eq:24}) represents that of a monochromatic wave of frequency $\nu$ whose amplitude $A(z,t)$ varies periodically with a much smaller frequency $\Delta \nu$ (cf. beat phenomenon). As an exercise, it is instructive to set $a(\nu')$ constant in Eq.~($\ref{eq:25}$) and establish that indeed $A(z,t)$ varies as a function of time with a frequency $\Delta \nu$. This modulation therefore effectively splits the monochromatic wave trains having different but nearly similar frequencies into wave groups whose length is of the order of $\lambda^2 / \Delta \lambda$, with $\Delta \lambda = -c \Delta \nu / \nu^2$ and the frequency of the order $\Delta \nu$ (see Figure $\ref{Fig_21_final}$).

\subsection{Visibility of the interference fringes}

What becomes the visibility of the interference fringes in the Young's hole experiment for the case of a quasi-monochromatic source having a finite dimension?

\begin{figure}[h]
\sidecaption
\centering
\includegraphics[width=9cm]{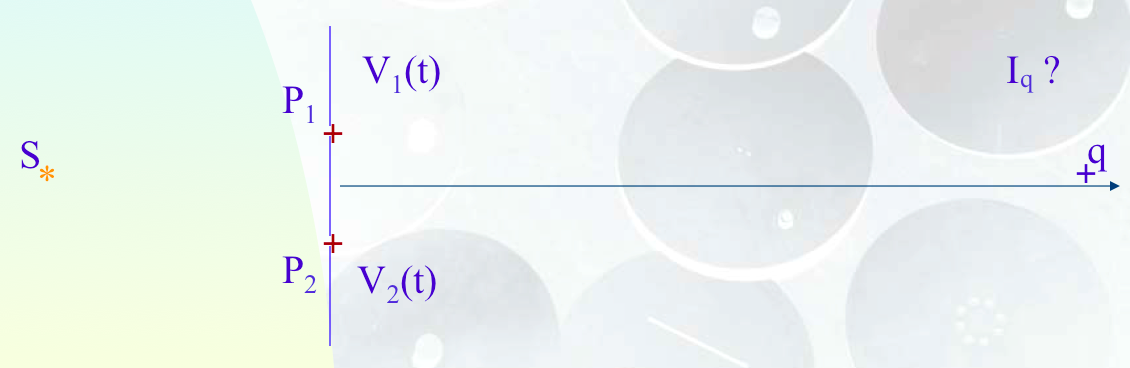}
\caption{Assuming an extended source S which quasi monochromatic light passes through the two holes $P_1$ and $P_2$, $I_q$ represents the intensity distribution at the point $q$ which accounts for the formation of the interference fringes. } 
\label{Fig_22_final}   
\end{figure}

   We can re-write the expression of the intensity $I_q$ at point $q$ as indicated below (see Eqs.~(\ref{eq:26})-(\ref{eq:29})). It is assumed that the holes placed at the points $P_1$, $P_2$ in the Young plane have the same aperture size (i.e. $V_1(t)$ = $V_2(t)$) and that the propagation times of the light between $P_1$ (resp. $P_2$) and $q$ are $t_{1q}$ (resp. $t_{2q}$, see Fig.~\ref{Fig_22_final}) :

\begin{equation}
\label{eq:26}
I_q = \left\langle V^\ast_q(t) V_q(t) \right\rangle,
\end{equation}

\begin{equation}
\label{eq:27}
V_q(t) = V_1(t - t_{1q}) + V_2(t - t_{2q})
\end{equation}

and after a mere change of the time origin

\begin{equation}
\label{eq:28}
V_q(t) = V_1(t) + V_2(t - \tau)
\end{equation}

where we have defined

\begin{equation}
\label{eq:29}
\tau = t_{2q} - t_{1q}.
\end{equation}

It follows that Eq.~(\ref{eq:26}) can be easily transformed into (\ref{eq:30}) where (\ref{eq:31}) represents the complex degree of mutual coherence, and the intensity $I = \thinspace  <V_1 V_1^\ast>  \thinspace = \thinspace <V_2 V_2^\ast>$. Equation (\ref{eq:30}) is used to find what is the intensity distribution of the interference fringes in the observation plane. The complex degree of mutual coherence $\gamma_{12}(\tau)$  (see Eq.~(\ref{eq:31})) is a fundamental quantity whose significance will be highlighted when calculating the visibility of the interference fringes. By means of (\ref{eq:24}), this function $\gamma_{12}(\tau)$  can still be expressed as (\ref{eq:32}), and if $ \tau << 1/\Delta\nu$  (i.e. the difference between the arrival times of the two light rays is less than the beat period $1/\Delta \nu$  of the quasi-monochromatic radiation), we can give it the form (\ref{eq:33}):

\begin{equation}
\label{eq:30}
I_q = I + I + 2I \thinspace Re[ \gamma_{12}(\tau) ],
\end{equation}

\begin{equation}
\label{eq:31}
\gamma_{12}(\tau) = \left\langle V_1^\ast(t) V_2(t-\tau) \right\rangle / I,
\end{equation}

\begin{equation}
\label{eq:32}
\gamma_{12}(\tau) = \left\langle A_1^\ast(z,t) A_2(z,t-\tau) \right\rangle \thinspace \exp(-i2\pi\nu\tau) / I,
\end{equation}

and if $\tau << 1/\Delta \nu$

\begin{equation}
\label{eq:33}
\gamma_{12}(\tau) = |\gamma_{12}(\tau=0)|  \thinspace \exp(i\beta_{12}-i2\pi\nu\tau).
\end{equation}

Equation (\ref{eq:30}) can then be rewritten as (\ref{eq:34}) and in this case the visibility $v$ of the interference fringes is $|\gamma_{12}(\tau=0)|$ (see Eq.~(\ref{eq:35})), $I_{max}$ and $I_{min}$ representing the brightest and weakest fringe intensities.

\begin{equation}
\label{eq:34}
I_q = I + I + 2 I \thinspace |\gamma_{12}(\tau=0)| \thinspace cos(\beta_{12}-2\pi\nu\tau)
\end{equation}
and 

\begin{equation}
\label{eq:35}
v=\left(\dfrac{I_{\max}-I_{\min}}{I_{\max}+I_{\min}}\right) = |\gamma_{12}(\tau=0)|.
\end{equation}

We will see in the next section that the module of $\gamma_{12}(\tau=0)$  is directly related to the structure of the source that we are observing. 

   We propose hereafter to the reader to answer the two following questions. What is the value of $|\gamma_{12}(\tau=0)|$ in the Young's holes experiment for the case of a monochromatic wave, two point-like holes and an infinitely small point-like source? And what can we say about the source when $|\gamma_{12}(\tau=0)| = 0$? 

   Let us now evaluate what $\gamma_{12}(\tau=0)$ is for the case we are interested in, namely an extended source emitting quasi-monochromatic light. This leads us directly to study the notion of the spatial coherence of light.

\subsection{Spatial coherence}

Let us thus evaluate Eq.~(\ref{eq:31}) for the case $\tau=0$. We find

\begin{equation}
\label{eq:36}
\gamma_{12}(\tau=0)=\left\langle V_{1}^\ast(t) 
V_{2}(t) \right\rangle / I.
\end{equation}

\begin{figure}[h]
\sidecaption
\centering
\includegraphics[width=9cm]{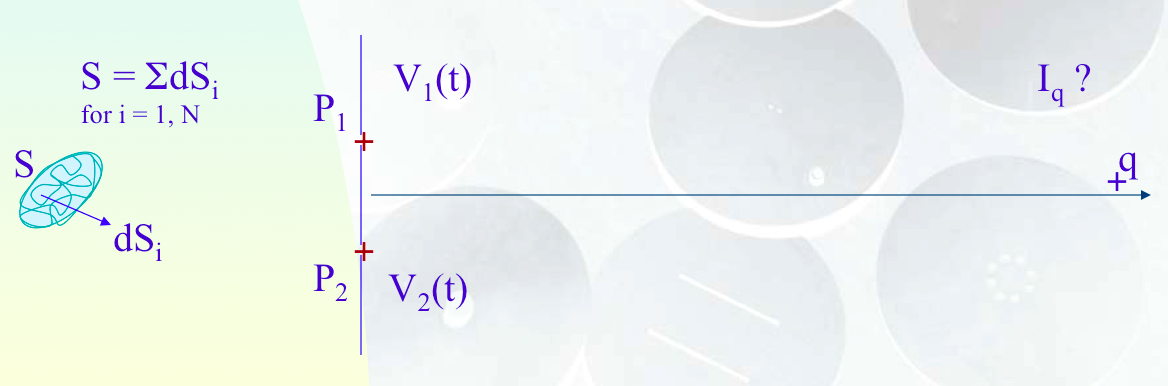}
\caption{The extended source $S$ is assumed to be composed of a large number of infinitesimal surface elements $dS_i$.} 
\label{Fig_23_final}   
\end{figure}

If $V_{i1}(t)$ and $V_{i2}(t)$ represent the electric fields at $P_1$ and $P_2$ due to a small surface element $dS_i$ on the source $S$ (see Fig.~\ref{Fig_23_final}), we find that the fields $V_1(t)$ and $V_2(t)$ can be expressed as


\begin{equation}
\label{eq:37}
\begin{split}
V_1(t) = \sum_{i=1}^N V_{i1}(t),  \\ 
V_2(t) = \sum_{i=1}^N V_{i2}(t). 
\end{split}
\end{equation}

It is assumed that the distinct points $i$ of the source are separated by small distances compared to the wavelength $\lambda$ of the light they emit in a mutually incoherent manner. Obtaining the expression (\ref{eq:38}) for $\gamma_{12}(0)$ is then immediate

\begin{equation}
\label{eq:38}
\gamma_{12}(0)=\left[\sum_{i=1}^N \left\langle V_{i1}^\ast
V_{i2} \right\rangle +\sum_{i\neq j}^N \langle V_{i1}^\ast
V_{j2}\rangle \right] / I.
\end{equation}

For an incoherent light source, the second summation appearing in (\ref{eq:38}) is obviously equal to zero.
   As a reminder, the contributions $V_{ij}(t)$ can be expressed as 

\begin{equation}
\label{eq:39}
\begin{split}
V_{i1}(t) = \frac{a_i(t-r_{i1}/c)}{r_{i1}} \thinspace \exp[i2\pi\nu (t-r_{i1}/c)], \hspace{0.5cm} \\
V_{i2}(t) = \frac{a_i(t-r_{i2}/c)}{r_{i2}} \thinspace \exp[i2\pi\nu (t-r_{i2}/c)]
\end{split}
\end{equation}

where $r_{i1}$ and $r_{i2}$ respectively represent the distances between the element $i$ of the source and the points $P_1$ and $P_2$. The products $V_{i1}^\ast(t) V_{i2}(t)$ simplify themselves as 

\begin{equation}
\label{eq:40}
V^{\ast}_{i1}(t) V_{i2}(t)= \frac{|a_i(t-r_{i1}/c)|^2}{r_{i1} r_{i2}} \thinspace \exp[-i2\pi\nu (r_{i2}-r_{i1})/c)],
\end{equation}

as long as the following condition is verified

\begin{equation}
\label{eq:41}
\left|r_{i1}-r_{i2}\right|\leq c / \Delta \nu=\lambda^2 /
\Delta \lambda=\ell.
\end{equation}

We thus see how to naturally introduce  the coherence length $\ell$ which characterizes the precision with which we must obtain the equality between the optical paths in order to be able to observe interference fringes (typically 2.5 microns in the visible for $\Delta \lambda$ = 1000 $\AA$).

\subsection{Zernicke-van Cittert theorem}

To obtain the mutual intensity due to the whole source, it suffices to insert in the expression (\ref{eq:38}), the relation (\ref{eq:40}) using (\ref{eq:42}). The result is Eq.~(\ref{eq:43}), also known as the Zernicke-van Cittert Theorem

\begin{equation}
\label{eq:42}
I(s)ds = |a_i(t-r/c)|^2,
\end{equation}

\begin{equation}
\label{eq:43}
\gamma_{12}(0) = \int_S \frac{I(s)}{r_1 r_2} \thinspace \exp[-i2\pi (r_{2}-r_{1})/\lambda)] ds / I.
\end{equation}

\begin{figure}[h]
\sidecaption
\centering
\includegraphics[width=9cm]{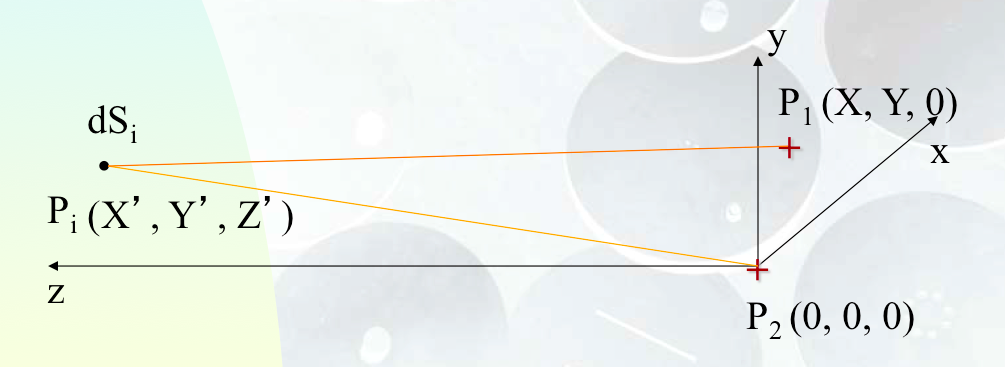}
\caption{Positions of the two elements $P_1$ and $P_2$ of the interferometer and of the infinitesimal element $P_i$ of the source assuming that the distance $Z' >> |X'|, |Y'|, |X|$ or $ |Y|$. } 
\label{Fig_24_final}   
\end{figure}

When the distance between the source and the screen is very large, the expression of this theorem can be simplified as follows. Let us adopt the orthonormal coordinate system ($x$, $y$, $z$) shown in Fig.~\ref{Fig_24_final} such that the coordinates of the two elements $P_1$ and $P_2$ of the interferometer are respectively ($X$, $Y$, $0$) and ($0$, $0$, $0$) and those of an infinitesimal element $dS_i$ of the source ($X'$, $Y'$ , $Z'$). It is then easy to find, by means of a relation analogous to (\ref{eq:16}), that

\begin{equation}
\label{eq:44}
|r_2-r_1| = |P_2P_i-P_1P_i| = |-\frac{(X^2+Y^2)}{2Z'}+(X \zeta + Y \eta)|
\end{equation}
where

\begin{equation}
\label{eq:45}
\zeta = \frac{X'}{Z'}, \thinspace \eta = \frac{Y'}{Z'}
\end{equation}
represent the angular coordinates of the source measured from the interferometer. Using the two last relations, one can easily transform the expression (\ref{eq:43}) into (\ref{eq:46}). The $X$, $Y$ coordinates in the first member of $\gamma_{12}(0, X/\lambda, Y/\lambda)$  represent the position of one element of the interferometer relative to the other. One often defines $u = X/ \lambda$ and $v = Y/\lambda$ which are quantities having the dimensions of the inverse of an angle, thus of angular space frequencies.

Apart from a multiplicative factor, we thus find that the visibility of the fringes (the function $|\gamma_{12}(\tau=0)|$) is simply the modulus of the Fourier transform of the normalized surface brightness $I'$ of the source (Eq.~(\ref{eq:47})). 

\begin{equation}
\label{eq:46}
\gamma_{12}(0, X/\lambda, Y/\lambda) = \thinspace \exp(-i\phi_{X,Y}) \int\int_S I'(\zeta, \eta) \thinspace \exp[-i2\pi(X\zeta+Y\eta)/\lambda]d\zeta d\eta 
\end{equation}

with 

\begin{equation}
\label{eq:47}
I'(\zeta, \eta) = \frac{I(\zeta, \eta)}{ \int\int_S I(\zeta', \eta') d\zeta' d\eta'}.
\end{equation}

In terms of the angular space frequencies $u = X/\lambda$, $v = Y/\lambda$, Eq.~(\ref{eq:46}) becomes

\begin{equation}
\label{eq:48}
\gamma_{12}(0, u, v) = \thinspace \exp(-i\phi_{u,v}) \int\int_S I'(\zeta, \eta) \thinspace \exp[-i2\pi(u\zeta+v\eta)]d\zeta d\eta.
\end{equation}

By a simple inverse Fourier transform, it is then possible to recover the (normalized) surface brightness of the source with an angular resolution equivalent to that of a telescope whose effective diameter would be equal to the baseline of the interferometer consisting of two independent telescopes 

\begin{equation}
\label{eq:49}
I'(\zeta, \eta) = \int\int \gamma_{12}(0, u, v) \thinspace \exp(i\phi_{u,v}) \thinspace \exp[i2\pi(\zeta u+ \eta v)]du dv.
\end{equation}

Equations (\ref{eq:48}) and (\ref{eq:49}) thus clearly highlight the power of the complex degree of mutual coherence since they make it possible to link the visibility and the normalized intensity distribution of the source by means of the Fourier transform $v = |\gamma_{12}(0)| = |FT[I']|$, and its inverse. Aperture synthesis consists in observing a maximum number of visibilities of the source, thus trying to cover as well as possible the $(u, v)$ plane from which we shall try, sometimes with some additional assumptions, to determine the structure of the source from the inverse Fourier transform (\ref{eq:49}) in which the integrant is not the visibility (i.e. the module of the complex degree of mutual coherence) but the complex degree of mutual coherence itself, within the factor $exp(i\phi_{x,y}$). 
   It is now good to remind some specific properties of the Fourier transform.

\subsection{Some remarkable properties of the Fourier transform and applications}

Let us remind that the Fourier transform of the function $f(x)$, denoted $FT[f(x)](s)$, where $x \in \Re$, is the function

\begin{equation}
\label{eq:50}
FT[f(x)](s)=\int_{-\infty}^\infty f(x) \thinspace \exp({-2i\pi s x }) dx
\end{equation}

where $s \in \Re$. The functions $f$ and $FT[f]$ form a Fourier pair. The function $FT[f]$ exists if the function $f(x)$ is bounded, summable and has a finite number of extrema and discontinuities. This does not necessarily imply that the inverse Fourier transform, denoted $FT^{-1}[FT[f]]$ transform is $f$. For the Fourier transformation to be reciprocal,

\begin{equation}
\label{eq:51}
f(x) =\int_{-\infty}^\infty FT[f](s) \thinspace \exp({2i\pi x s }) ds,
\end{equation}

it suffices $f$ to be of summable square, i.e. that the following integral exists

\begin{equation}
\label{eq:52}
\int_{-\infty}^\infty |f(x)|^2   dx.
\end{equation}

   The definition of $FT$ can be extended to the distributions. The $FT$ of a distribution is not necessarily of summable square. Let us also note that the functions $f$ and $FT[f]$ can be real or complex.

   We can generalize the $FT$ to several dimensions, by defining $f$ on the space $\Re^n$. Let $\bf{r}$, $\bf{w} \in \Re^n$, we then have
   
\begin{equation}
\label{eq:53}
FT[f]({\bf w})=\int_{-\infty}^\infty f({\bf r}) \thinspace \exp({-2i\pi {\bf w} {\bf r} }) d{\bf r}.
\end{equation}

As a reminder, if $f(t)$ designates a function of time, $FT[f](s)$ represents its content as a function of time frequencies. Similarly, if $f(\bf{r})$ is defined on $\Re^2$, where $\Re^2$ represents a two-dimensional space, the function $FT[f](\bf{w})$ represents the space frequency content of $f(\bf{r})$, where $\bf{w} \in \Re^2$.

   Among the interesting properties of the Fourier transform, let us remind:

\subsubsection{Linearity:}

\begin{equation}
\label{eq:54}
FT[af] = a \thinspace FT[f],
\end{equation}

with the constant $a \in \Re$,
 
\begin{equation}
\label{eq:55}
FT[f + g] =  FT[f] + FT[g].
\end{equation}

\subsubsection{Symmetry and parity:}

The considerations of symmetry are very useful during the study of the Fourier transform. Let $P(x)$ and $I(x)$ be the even and odd parts of $f(x)$ such that

\begin{equation}
\label{eq:56}
f(x) = P(x) + I(x),
\end{equation}

we find that

\begin{equation}
\label{eq:57}
FT[f](s)=2\int_0^\infty P(x)\cos (2\pi xs)dx-2i
\int_0^\infty I(x)\sin (2\pi xs)dx.
\end{equation}

From this result, we can deduce for instance that if $f(x)$ is real, the real part of $FT[f](s)$ will be even and its imaginary part will be odd whereas if $f(x)$ is complex, the imaginary part of $FT[f](s)$ will be even and its real part will be odd. 

\subsubsection{Similarity:}
The relationship of similarity is the following one

\begin{equation}
\label{eq:58}
FT[f(x/a)](s) = |a| \thinspace FT[f(x)](as)
\end{equation}

where $a \in \Re$, is a constant. The dilation of a function causes a contraction of its Fourier transform. This very visual property is very useful to understand that a function whose support is very compact, has a very spread transform. In the analysis of temporal frequencies, one would state that a pulse of very short duration results in a very broad frequency spectrum, that is to say, contains frequencies all the higher as the pulse is brief. This is the classical relation of the spectrum of a wave packet, according to which the knowledge of the properties of a signal cannot be arbitrarily precise both in time and in frequency.

\subsubsection{Translation:}

The translation relation is written as

\begin{equation}
\label{eq:59}
FT[f(x-a)](s) = \exp({-2i\pi as}) \thinspace FT[f(x)](s).
\end{equation}

A translation of the function in its original space corresponds to a phase rotation of its Fourier transform in the transformed space.

\subsubsection{Door function:}

\begin{figure}[h]
\sidecaption
\centering
\includegraphics[width=9cm]{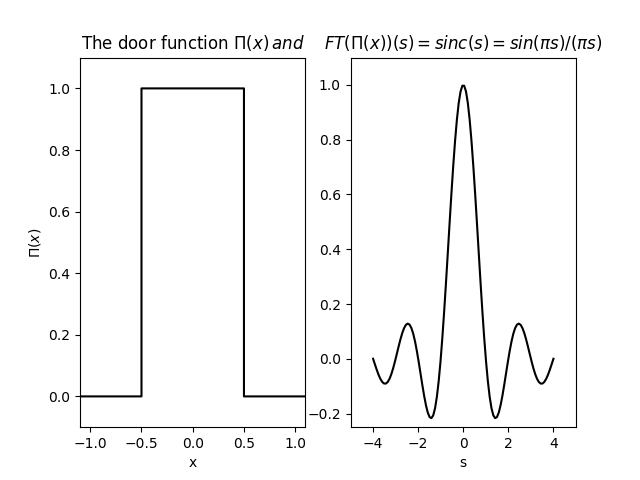}
\caption{The door function and its Fourier transform (cardinal sine).} 
\label{Fig_25_final}   
\end{figure}

The door function, denoted $\Pi(x)$, is defined by (see Fig.~ \ref{Fig_25_final})

\begin{equation}
\label{eq:60}
\Pi(x)=1 \quad {\rm if}\;  x \in [-0.5, 0.5],  \quad {\rm and} \quad \Pi(x)=0 \quad {\rm otherwise.}
%
\end{equation}

It is easy to find that

\begin{equation}
\label{eq:61}
FT[\Pi(x)](s) = sinc(s) = \frac{sin(\pi s)}{\pi s}.
\end{equation}

Applying the similarity relation, we also find that

\begin{equation}
\label{eq:62}
FT[\Pi(x/a)](s) = |a| \thinspace sinc(as) = |a| \thinspace \frac{sin(\pi as)}{\pi as}. 
\end{equation}

The door function is also sometimes called the window function or simply window.

\subsubsection{Distribution of Dirac:}

The Dirac distribution, also called Dirac peak, is noted $\delta(x)$. It is defined by the following integral, which exists only in the sense of the distributions

\begin{equation}
\label{eq:63}
\delta(x) = \int_{-\infty}^\infty \exp({2i\pi xs}) ds.
\end{equation}

Its Fourier transform is therefore 1 in the interval $\left] -\infty, +\infty \right[$  since $\delta(x)$ appears above as the inverse Fourier transform of 1.

\subsubsection{Applications:}

We propose hereafter several astrophysical applications that make use of the previous remarkable properties of the Fourier transform. 

   Let us first consider the case of a double star which two point-like components are equally bright and separated by an angle  2$\zeta_0$. Making use of Eqs.~(\ref{eq:35}), (\ref{eq:47}) and (\ref{eq:48}), one may easily establish using the properties (\ref{eq:59}) and (\ref{eq:63}) that the normalized intensity $I'(\zeta)$ takes the simple form

\begin{equation}
\label{eq:64}
I'(\zeta) = \frac{\delta(\zeta - \zeta_0) + \delta(\zeta + \zeta_0)}{2}
\end{equation}

and that the visibility $v$ measured with an interferometer composed of 2 telescopes separated by the baseline $X$ is given by the expression

\begin{equation}
\label{eq:65}
v = |\gamma_{12}(0)| = |cos(2 \pi \zeta_{0} u)|
\end{equation}

where $u = X / \lambda$.

A second nice application consists in deriving the visibility of the interference fringes measured with the same interferometer of a $1{-}D$ Gaussian star which intensity $I(\zeta)$ distribution is given by the following expression 

\begin{equation}
\label{eq:66}
I(\zeta) = \thinspace \exp (\dfrac{-4 \ln (2) \zeta^2}{\hbox{FWHM}^2})
\end{equation}

where $FWHM$ represents the angular full width at half maximum of the $1{-}D$ Gaussian star. The expression of the corresponding visibility is then easily found to be

\begin{equation}
\label{eq:67}
v= \left|\gamma_{12}(0,u)\right|=\thinspace \exp \left(\dfrac{-\pi^2 u^2\
\hbox{FWHM}^2}{4 \ln (2)}\right),
\end{equation}

and we notice that narrower is the angular size of the star, broader is its visibility content in angular space frequencies.  

   In the third proposed application, we ask to establish the expression of the visibility of a $2{-}D$ uniformly bright square star which each angular side is $\zeta_0$,  i.e. $I(\zeta) = Cte ~ \Pi(\zeta / \zeta_0) ~ \Pi(\eta / \zeta_0)$.

   The expression to be derived is the following one

\begin{equation}
\label{eq:68}
v = |\gamma_{12}(0,u,v)| = |\frac{sin(\pi \zeta_0 u)}{\pi \zeta_0 u} \frac{sin(\pi \zeta_0 v)}{\pi \zeta_0 v}|.
\end{equation}

Finally, a generalization of the previous application consists in deriving the visibility of a star which is seen as a projected $2{-}D$ uniform circular disk which angular radius is $\rho_{UD}$ and its angular diameter $\theta_{UD}$.

Due to the circular symmetry of the problem, it is convenient to make use of polar coordinates in Eq.~(\ref{eq:46}) as follows:

\begin{equation}
\label{eq:2000}
\begin{split}
u = X / \lambda = R \cos(\psi) / \lambda  \\
v = Y / \lambda = R \sin(\psi) / \lambda, 
\end{split}
\end{equation}

where $R$ denotes the baseline between the two telescopes of the interferometer, and

\begin{equation}
\label{eq:2001}
\begin{split}
\zeta = \theta \cos(\phi)  \\
\eta = \theta \sin(\phi).
\end{split}
\end{equation}

Eq.~(\ref{eq:46}) then transforms into

\begin{equation}
\label{eq:2002}
|\gamma_{12}(0,R / \lambda, \psi)|=|\dfrac{1}{\pi \rho_{UD}^2} \int_{0}^{\rho_{UD}} \theta \int_{0}^{2 \pi} \exp{[-i2\pi \theta R / \lambda (\cos(\phi) \cos(\psi) + \sin(\phi) \sin(\psi))]} d\phi d\theta|.
\end{equation}

Making use of the additional changes of variables

\begin{equation}
\label{eq:2003}
\begin{split}
z = 2 \pi \theta R / \lambda  \\
\Phi = \pi / 2 - \phi + \psi,
\end{split}
\end{equation}

Eq.~(\ref{eq:2002}) becomes

\begin{equation}
\label{eq:2004}
|\gamma_{12}(0,R / \lambda, \psi)|=|{(\dfrac{\lambda}{2 \pi R})}^2 \dfrac{1}{\pi \rho_{UD}^2} \int_{0}^{2 \pi \rho_{UD} R / \lambda} \theta \int_{-3 \pi / 2 + \psi}^{\pi / 2 + \psi} \cos(z \sin(\Phi))  d\Phi d\theta|.
\end{equation}

Reminding the definition of the zero order Bessel function $J_0(x)$
\begin{equation}
\label{eq:2005}
J_0(x) = \dfrac{1}{\pi} \int_{0}^{\pi} \cos[x  \sin(\theta)] d\theta,
\end{equation}
and the relation existing between $J_0(x)$ and the first order Bessel function $J_1(x)$, namely

\begin{equation}
\label{eq:2006}
x J_1(x) = \int x' J_0(x') dx', 
\end{equation}

\begin{figure}[h]
\sidecaption
\centering
\includegraphics[width=9cm]{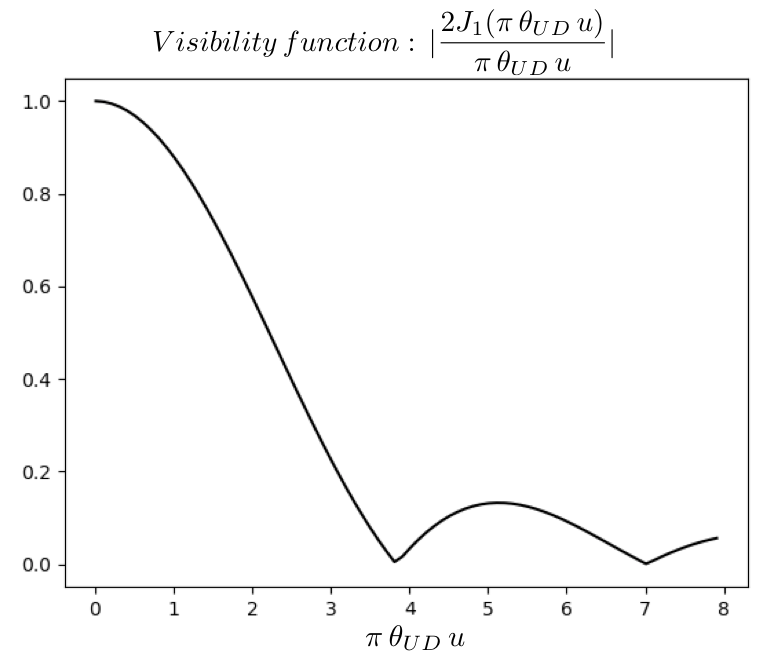}
\caption{Visibility function expected for a star consisting of a uniformly bright circular disk with an angular diameter $\theta_{UD}$. } 
\label{Fig_26_final}   
\end{figure}

Eq.~(\ref{eq:2004}) successively reduces to 

\begin{equation}
\label{eq:2007}
|\gamma_{12}(0,R / \lambda)|=|{(\dfrac{\lambda}{2 \pi R})}^2 \dfrac{1}{\pi \rho_{UD}^2} 2 \pi \int_{0}^{2 \pi \rho_{UD} R / \lambda} z J_0(z) dz| 
\end{equation}
and
\begin{equation}
\label{eq:2008}
|\gamma_{12}(0,R / \lambda)|= |2 \dfrac{J_1(2 \pi \rho_{UD} R / \lambda)}{2 \pi \rho_{UD} R / \lambda}|.  
\end{equation}
We thus find that the expression (\ref{eq:35}) of the fringe visibility for the case of a star seen as a projected $2{-}D$ uniform circular disk with an angular dimater $\theta_{UD} = 2 \rho_{UD}$ is

\begin{equation}
\label{eq:69}
v=\left(\dfrac{I_{\max}-I_{\min}}{I_{\max}+I_{\min}}\right)=
|\gamma_{12}(0,u)|=\left|\dfrac{2J_1(\pi\theta_{UD}u)}
{\pi \theta_{UD}u}\right|,
\end{equation}

where we have set $u = R / \lambda$. As a reminder, the Bessel function has the following properties

\begin{equation}
\label{eq:2009}
\begin{split}
J_1( x = 3.8317...) = 0   \\
\lim_{x \to 0} \dfrac{J_1(x)}{x} = 1/2,
\end{split}
\end{equation}

which allow us to easily understand the behavior of the visibility function illustrated in Fig. \ref{Fig_26_final}.

\begin{figure}[h]
\sidecaption
\centering
\includegraphics[width=9cm]{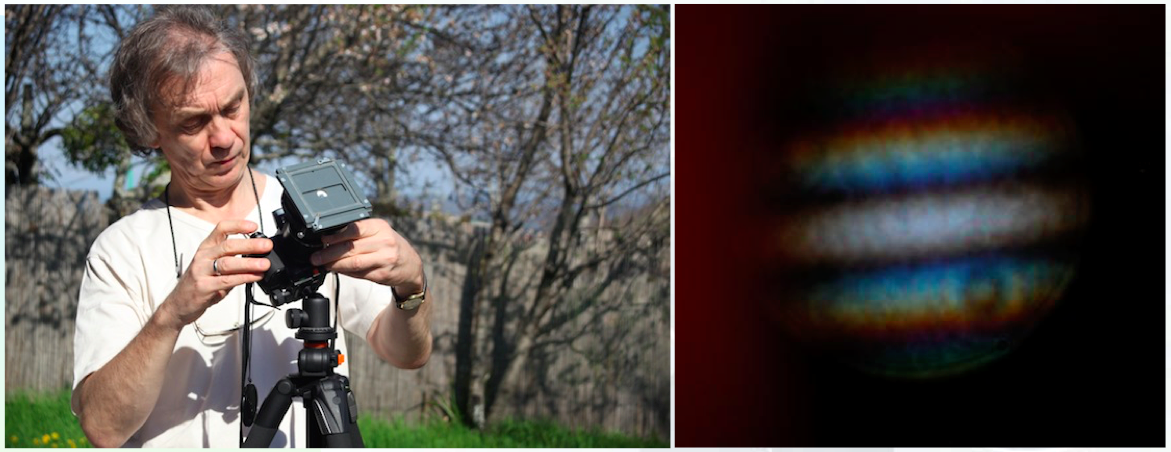}
\caption{Solar fringes photographed with an EOS 5D Canon camera in front of which was set a micro-interferometer consisting of two holes having a diameter of 11.8 $\mu$ separated by a baseline of 29.4 $\mu$.} 
\label{Fig_27_final}   
\end{figure}

One could then wonder whether it is possible to observe interferometric fringes from our nearest star, i.e. the Sun? Figure \ref{Fig_27_final} illustrates such fringes in while light obtained on 9th of April 2010 using a micro interferometer consisting of 2 holes with a diameter of 11.8 $\mu$ separated by a baseline of 29.4 $\mu$. This micro-interferometer was placed in front of the objective of an EOS 5D Canon camera. Since the picture was taken in white light, it is possible to see the effects due to color dispersion. It is then easy to get an estimate of the fringe visibility, using Eq.~(\ref{eq:69}), assuming that the Sun is a uniform disk with an angular diameter of 30'. 

\section{Some examples of interferometers }

\begin{figure}[h]
\sidecaption
\centering
\includegraphics[width=12cm]{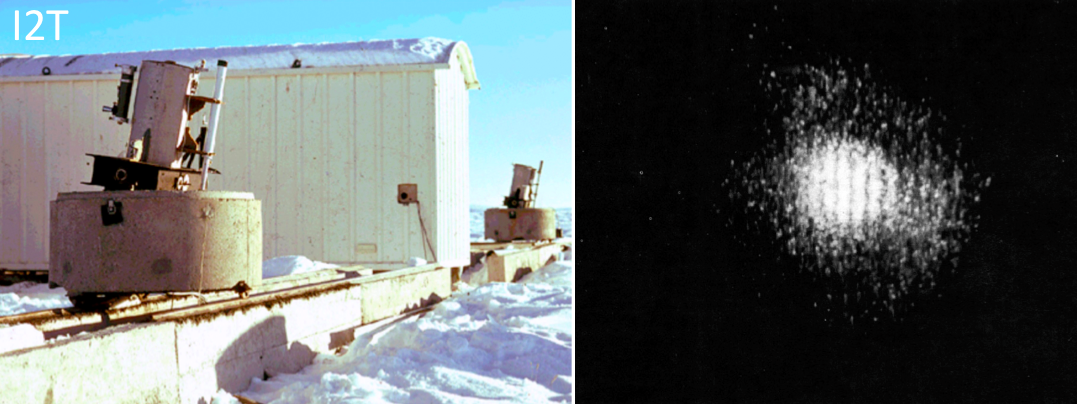}
\caption{First fringes obtained with the I2T on Vega (Labeyrie et al. 1975, \copyright \thinspace Observatoire de la C\^{o}te d'Azur ).} 
\label{Fig_28_final}   
\end{figure}

\begin{figure}[h]
\sidecaption
\centering
\includegraphics[width=9cm]{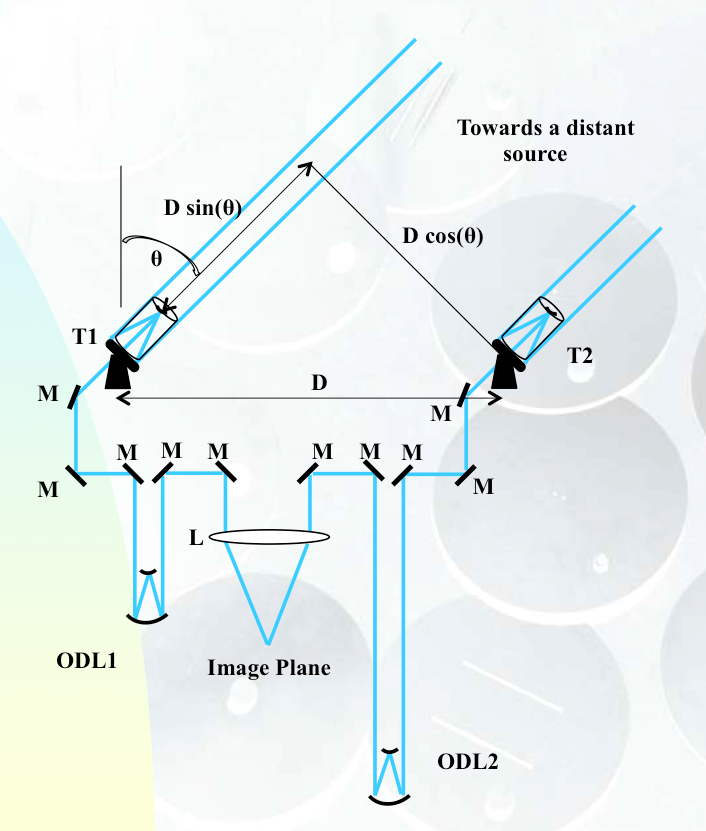}
\caption{Use of optical delay lines to compensate for the continuous change in the lengths of the two light paths as the Earth rotates. } 
\label{Fig_29_final}   
\end{figure}

\begin{figure}[h]
\sidecaption
\centering
\includegraphics[width=8cm]{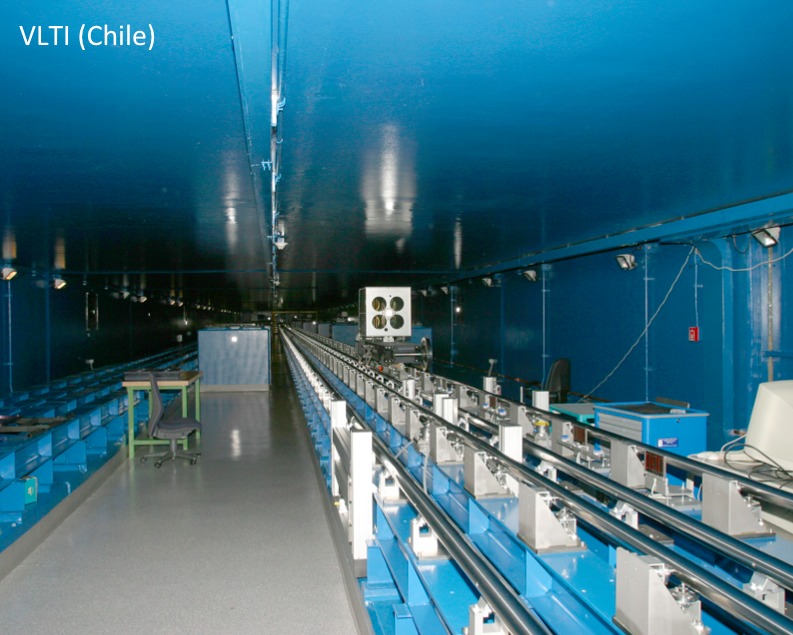}
\caption{View inside the optical delay line tunnel of the VLTI at ESO, Paranal, Chile. \copyright \thinspace ESO.} 
\label{Fig_7000_final}   
\end{figure}

\begin{figure}[h]
\sidecaption
\centering
\includegraphics[width=8cm]{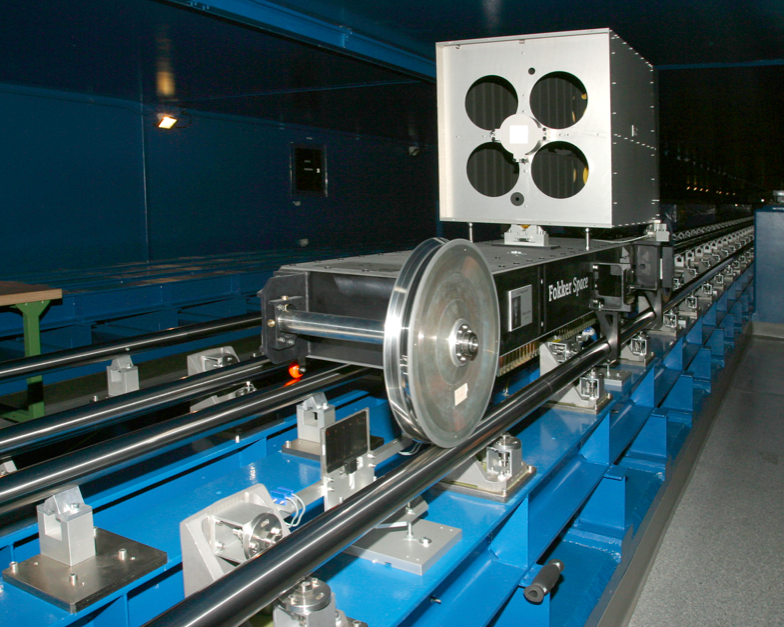}
\caption{Zoom on one of the optical delay lines used in the tunnel of the VLTI at ESO, Paranal, Chile. \copyright \thinspace ESO.} 
\label{Fig_7001_final}   
\end{figure}

One of the most respected sanctuaries of optical interferometry is located on the plateau of Caussols, north of Grasse, in the south of France. The I2T (in French, ''Interf\'{e}rom\`{e}tre \`{a} 2 T\'{e}lescopes''), made of 2 telescopes with an aperture of 26cm each and separated by a baseline of up to 144m was characterized by an angular resolution $\Phi \sim  0.001''$ attainable for objects with an apparent magnitude brighter than $V_{lim} \sim 6$ (see Figure \ref{Fig_28_final}, left). First interference fringes were obtained on Vega in 1975 (Fig.~\ref{Fig_28_final}, right). About twenty angular diameters of stars have been measured using the same I2T by Prof. Antoine Labeyrie and his close collaborators.
   In order to equalize the light paths collected from the stars passing through the two telescopes, optical delay lines are mandatory. These have been successfully used for the first time in 1975 (see Fig.~\ref{Fig_29_final} for an illustration of how delay lines work and Figs.~\ref{Fig_7000_final} and~\ref{Fig_7001_final} for views on modern optical delay lines in use at the VLTI, ESO, Chile).

\begin{figure}[h]
\sidecaption
\centering
\includegraphics[width=9cm]{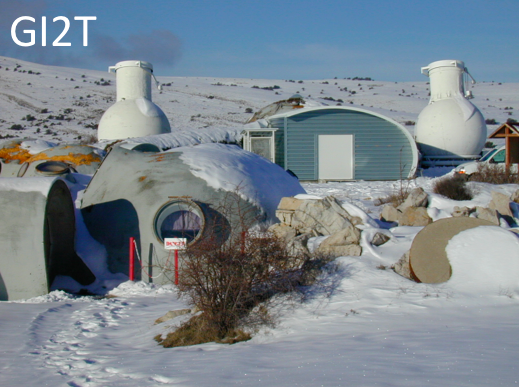}
\caption{The GI2T constructed by Antoine Labeyrie and his close collaborators on the plateau of Caussols, north of Grasse, near Nice (France, \copyright \thinspace Observatoire de la C\^{o}te d'Azur}). 
\label{Fig_30_final}   
\end{figure}

The GI2T (in French, ''Grand Interf\'{e}rom\`{e}tre \`{a} 2 T\'{e}lescopes'') composed of two 1.5m telescopes was subsequently used by the same team. The two big telescopes could in principle be set 2 km apart, corresponding to an angular resolution $\Phi \sim  0.0001''$  (see Fig.~\ref{Fig_30_final}). 

\begin{figure}[h]
\sidecaption
\centering
\includegraphics[width=9cm]{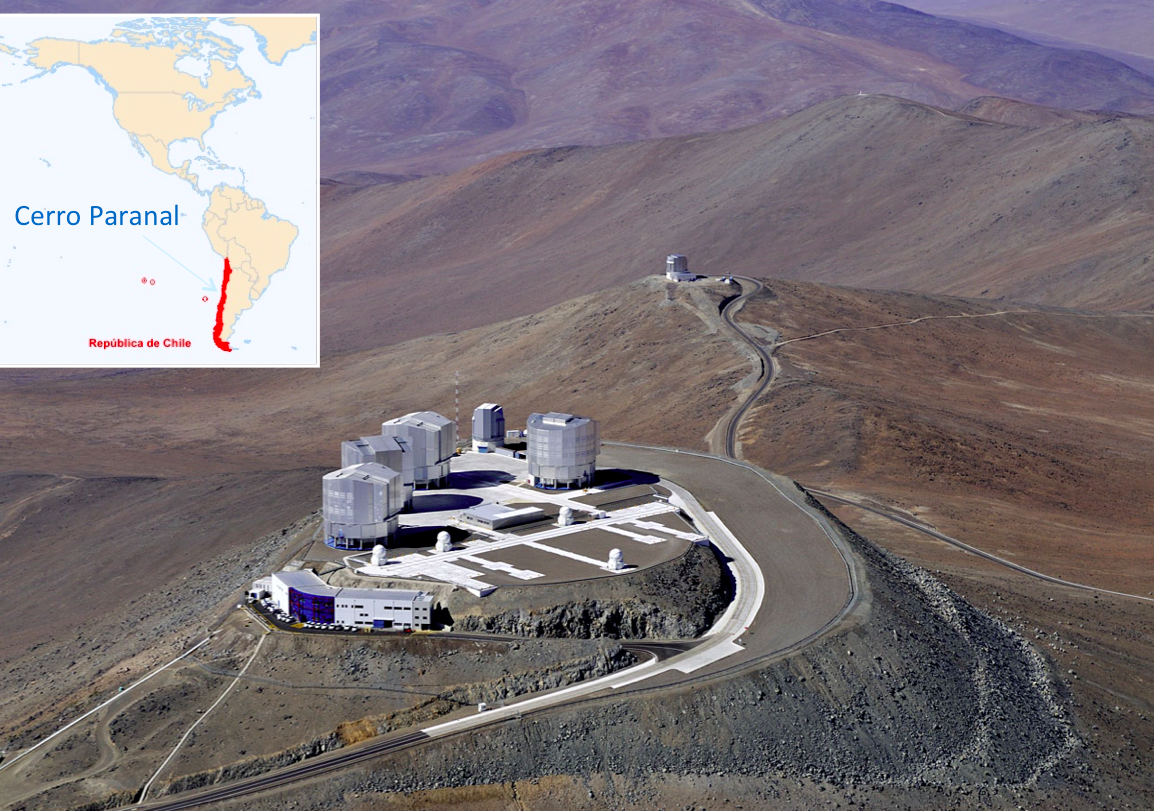}
\caption{The Very Large Telescope Interferometer (VLTI) at the top of Paranal (Chile, \copyright \thinspace ESO).} 
\label{Fig_31_final}   
\end{figure}

Since the beginning of the $21^{st}$ century, the modern sanctuary of stellar interferometry and aperture synthesis is undoubtedly the Very Large Telescope Interferometer (VLTI) of ESO (Southern European Observatory), located in Chile on Mount Paranal (see Fig.~\ref{Fig_31_final}).
   The VLTI is a European interferometer that can re-combine the signal from 2, 3 or 4 telescopes depending on the instrument used. It has 4 telescopes of 8.2m and 4 mobile telescopes of 1.8m. Only telescopes of the same size can be re-combined together. The auxiliary telescopes of 1.8m can be easily moved allowing a better coverage of the $u, v$ plane. The maximum base length of this interferometer is about 200m.

\begin{figure}[h]
\sidecaption
\centering
\includegraphics[width=9cm]{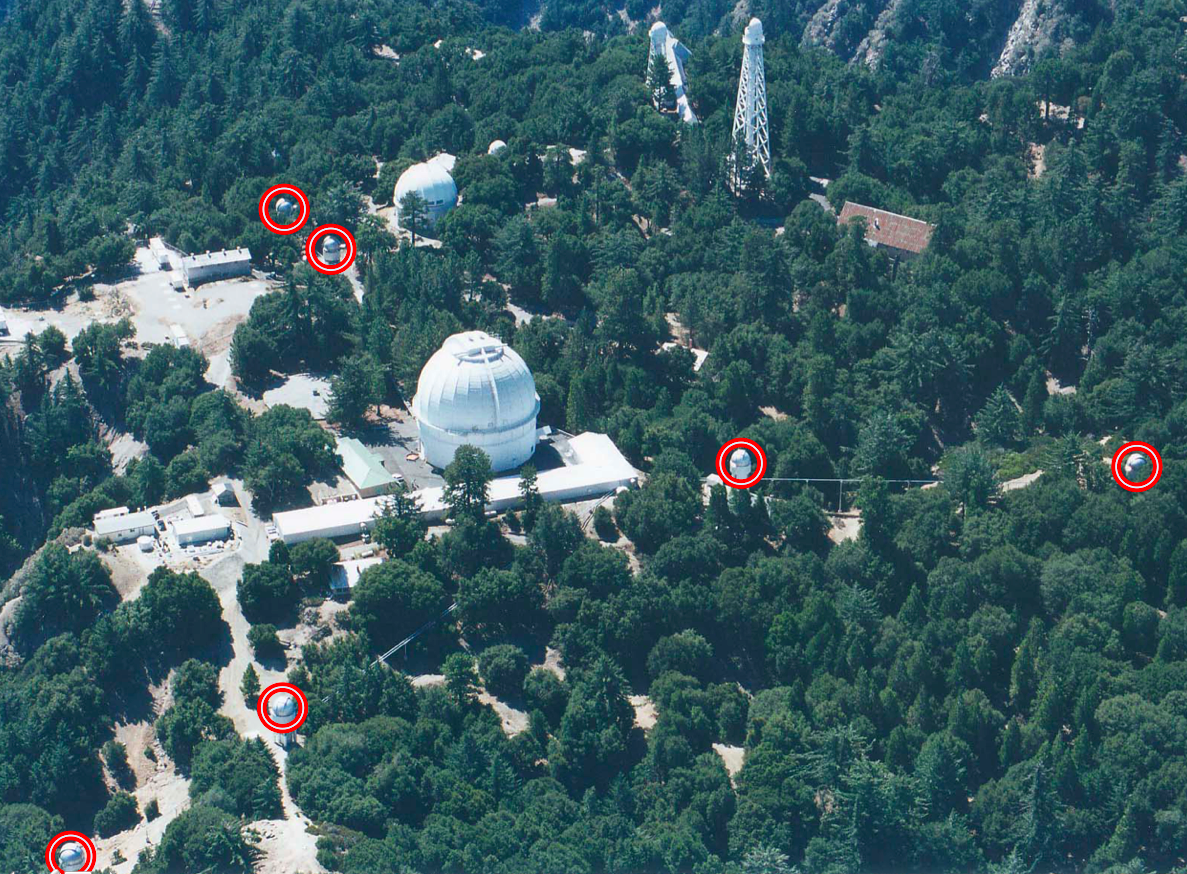}
\caption{The CHARA interferometer composed of six 1m telescopes at Mount Wilson Observatory (California, USA). \copyright \thinspace The Observatories of the Carnegie Institution.} 
\label{Fig_32_final}   
\end{figure}

CHARA is another very performing interferometer located on the heights of Los Angeles, California (see Fig.~\ref{Fig_32_final}). It is installed on the historic observatory of Mount Wilson. Remember that it was with the 2.5m telescope of this observatory that the first measurement of a stellar diameter was made by Michelson and Pease by installing a beam of 7m at the top of the telescope. The CHARA interferometric array, operational since 1999 is composed of 6 telescopes of 1m in diameter. These 6 telescopes can be either re-combined by 2, by 3 since 2008 and recently the 6 together. The maximum base length of this interferometer is 330m allowing to achieve an angular resolution of 200 $\mu$arcsec.

\begin{figure}[h]
\sidecaption
\centering
\includegraphics[width=9cm]{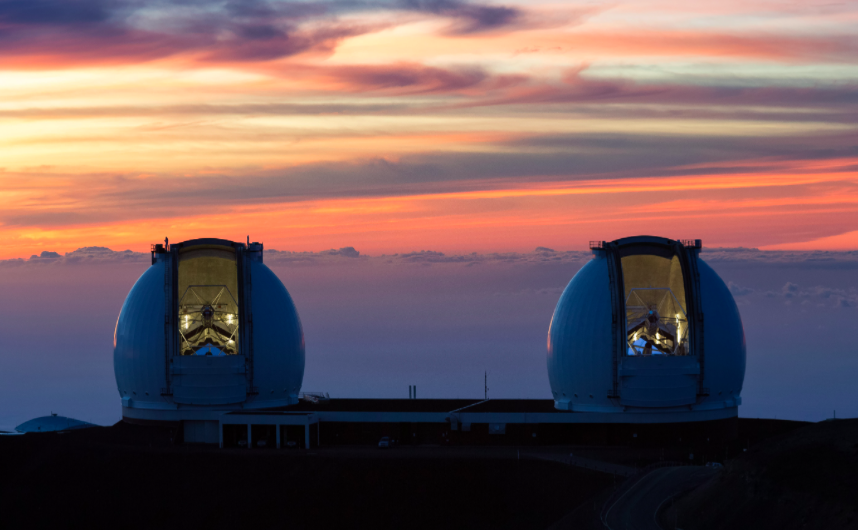}
\caption{The Keck interferometer on top of Mauna Kea (Hawaii, USA). \copyright \thinspace Ethan Tweedie.} 
\label{Fig_33_final}   
\end{figure}

It is mainly used for angular diameter measurements but also for the detection and characterization of tight binary stars as well as for the detection of exo-zodiacal clouds (clouds of dust gravitating around the stars).
   Another famous optical/IR interferometer is the Keck Interferometer made of two 10m telescopes separated by a fixed baseline of 85m (see Fig.~\ref{Fig_33_final}) on top of Mauna Kea (Hawaii, USA).

\section{Three important theorems and some applications}

When we previously established the relation existing between the structure of a celestial source and the visibility of the fringes observed with an interferometer (Sections 3.3 and 4.4), we implicitly assumed that the size of the apertures was infinitely small (pinhole apertures). Use of the fundamental theorem allows one to calculate the response function of an interferometer equipped with finite size apertures. This theorem actually formalizes, in mathematical terms, the physical connection existing between the focal plane and the pupil plane of an optical instrument (telescope, interferometer, grating, etc.). Use of the convolution theorem will then enable us to establish the relation between a celestial source that is extended and its observed image in the focal plane of an optical instrument. Finally, the Wiener-Khinchin theorem establishes the relation between the frequency content of the point spread function of an optical instrument and its pupil plane characteristics.     

\subsection{The fundamental theorem: relation between the pupil and focal planes}
 
The fundamental theorem that we shall demonstrate here merely stipulates that given a converging optical system which can be assimilated to the lens or to the mirror of a telescope, or of an optical interferometer, the complex amplitude distribution $a(p,q$) of the electromagnetic field of radiation in the focal plane is the Fourier transform of the complex amplitude distribution $A(x,y)$ of the electromagnetic field in the pupil plane, i.e.

\begin{equation}
\label{eq:70}
a(p,q) = \int_{R^2} A(x,y) \exp{[-i2\pi(px+qy)]} dx dy,
\end{equation}

or in a more compact form

\begin{equation}
\label{eq:71}
a(p,q) = FT[A(x,y)](p,q),
\end{equation}

with 

\begin{equation}
\label{eq:72}
p = \frac{x'}{\lambda f} \quad {\rm and} \quad q = \frac{y'}{\lambda f},
\end{equation}

where $x'$, $y'$ refer to the Cartesian coordinates in the focal plane, $\lambda$ to the wavelength of the monochromatic light under consideration and $f$ to the effective focal length of the converging system. 
\newline

\begin{figure}[h]
\centering
\includegraphics[width=9cm]{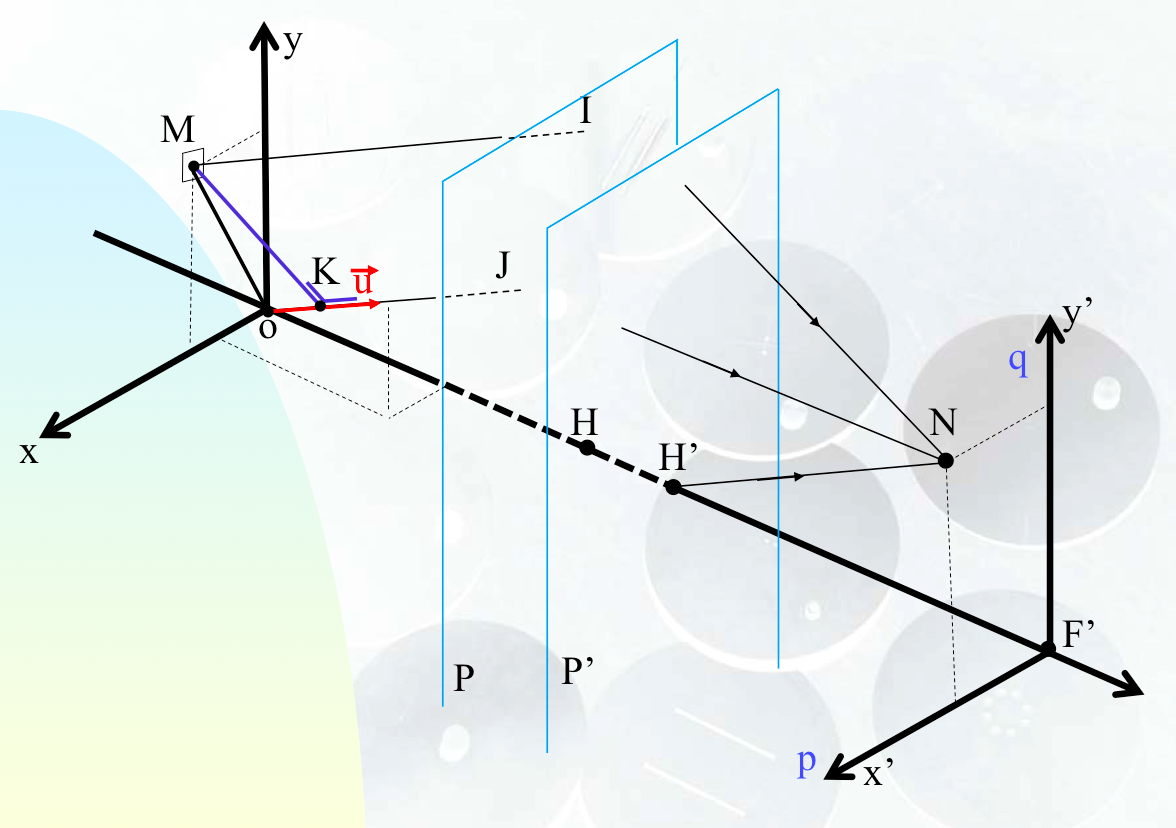}
\caption{Fourier transform by a focusing optical system represented by its main planes $P$ and $P'$. For the case of a thin lens, the latter would be degenerated into a single plane passing through its center.} 
\label{Fig_1000_final}   
\end{figure}

Figure~\ref{Fig_1000_final} represents a convergent optical system, its focal point $F'$, its principal planes $P$, $P'$ and its principal points $H$ and $H'$. The latter degenerate with the optical center in the case of a thin lens or with the bottom of the dish in the case of a single mirror.
The two orthonormal coordinate systems ($O$, $x$, $y$, $z$) and ($F'$, $x'$, $y'$, $z'$) make it possible to locate the input pupil plane and the image focal plane of the optical system. The term 'pupil plane' serves as the support for the definition of the vibration state at the entrance of the collector while the 'focal plane' serves as the support for the definition of the image that the collector gives of the source located at infinity. Defining the action of the collector is thus to establish the transformation that it operates on the radiation between these two planes. 

The {\bf hypotheses} underlying this theorem are: \newline
H1. The optical system is free from any geometric aberration. \newline
H2. The edges of the diaphragm do not disturb the electromagnetic field of radiation, that is to say that the diaphragm behaves as an ''all (1) or nothing (0)'' function with respect to this field. This is equivalent to assume that the dimensions of the collecting aperture(s) are large with respect to the wavelength of the light. \newline
H3. No disturbance, other than those imposed by the optical system, intervenes between the pupil and the focal planes. The optical elements are thus assumed to be perfectly transparent or reflective. \newline
H4. The light source is located at an infinite distance from the optical system and can thus be considered to be point-like. \newline
H5. The disturbances occurring between the source and the pupil plane are weak and have very long evolution times relative to the period (i.e. $T = 1/\nu = \lambda/c$) of the radiation. \newline
H6. The radiation is monochromatic and has a fixed polarization plane. \newline

{\bf Theorem statement:} \newline
Within a multiplicative coefficient of the variables, the amplitude distribution in the focal plane is the Fourier transform of the amplitude distribution in the pupil plane. 

\vspace{0.3cm}

{\bf Demonstration:} \newline
Consider the different points ($x$, $y$) of the pupil plane. H6 (i.e. the previous hypothesis 6) makes it possible to represent the electrical component of the electromagnetic field by the real part of the vibration distribution 
\begin{equation}
\label{eq:1001}
A(x,y) \exp{(i2 \pi \nu t)}, 
\end{equation}
with the very general representation of the expression of the complex amplitude $A(x,y)$
\begin{equation}
\label{eq:1002}
A(x,y) =  \mathcal{A}(x,y) \exp{(i \Phi(x,y))} P_0(x,y),
\end{equation}
where $\mathcal{A}(x,y)$ and $\Phi(x, y)$ represent the amplitude and phase of the electric field and $P_0(x,y)$ the input pupil function which is 1 inside the pupil and 0 outside (in agreement with H2 and H3).

In agreement with the Huygens-Fresnel principle, we will consider in the following that every point reached by a wave can be considered as a secondary source re-emitting a vibration with the same amplitude, the same frequency $\nu$, the same polarization and the same phase (within a constant phase shift of $\pi / 2$) as those of the incident vibration at this point.
The point $N$($x'$,$y'$) of the focal plane will thus receive vibrations emitted by all the points of the pupil plane. The laws of geometrical optics, deduced from the Fermat principle, make it possible to write that the rays which, after the optical system, converge at the point $N$ of the image focal plane, were, before the optical system, parallel to $H'N$. Having assumed that the source is at infinity (in agreement with H4), the amplitude will be preserved between the pupil plane and the focal plane.
From the point $M$($x$,$y$) of the pupil plane, the point $N$($x'$,$y'$) of the focal plane will thus receive the vibration
\begin{equation}
\label{eq:1003}
A(x,y) \exp{(i2 \pi \nu t + i \Psi)}. 
\end{equation}
Let us take as the zero phase shift reference that of the ray passing through the point $O$ along the direction $OJN$ . 
The phase shift $\Psi$ can then be expressed using the difference between the optical paths
\begin{equation}
\label{eq:1004}
\delta = d(M I N) - d(O J N), 
\end{equation}
where $d()$ refers to the distance along the specified path, and the relation
\begin{equation}
\label{eq:1005}
\Psi = 2 \pi \delta / \lambda. 
\end{equation}

If the point $K$ corresponds to the orthogonal projection of $M$ onto $OJ$, $M$ and $K$ belong to the same wave plane which, after the optical system, will converge at the $N$ point of the focal plane. \newline
The Fermat principle, according to which the optical path between a point and its image is constant (rigorous stigmatism) or extremum (approximate stigmatism) makes it possible to write that the difference in optical path ($MIN$)$-$($KJN$) behaves in the neighborhood of zero as an infinitely small second order with respect to the $d$($I$,$J$) and thus also with respect to $d$($O$,$M$) and $d$($O$,$K$), which are of the same order as $d$($I$,$J$). As a result (see Fig.~\ref{Fig_1000_final}),

\begin{equation}
\label{eq:1006}
\delta = -d(O,K) = -|(\bf{OM \thinspace u})|, 
\end{equation}

$\bf{u}$ designating the unit vector along the direction $H'N$ and $(\bf{OM \thinspace u})$ the scalar product between the vectors $\bf{OM}$ and $\bf{u}$.
If the angle that $H'N$ makes with the optical axis is small, the vector of components ($x'/f$, $y'/f$, $1$) is the vector director of $H'N$ and has a norm close to 1 (at first order because $f >> |x '|, | y' |$). Moreover, $\bf{OM}$ has for components ($x$, $y$, $0$). Using 
Eq.~(\ref{eq:1006}) in (\ref{eq:1005}), the expression (\ref{eq:1003}) becomes

\begin{equation}
\label{eq:1007}
A(x,y) \exp{(i2 \pi \nu t - x x' / \lambda f - y y' / \lambda f )}.
\end{equation}

Choosing as new variables in the focal plane those defined in (\ref{eq:72}), we get

\begin{equation}
\label{eq:1008}
A(x,y) \exp{(-i2 \pi ( x p + y q )} \exp{(i2 \pi \nu t)}.
\end{equation}

The resulting vibration at the point $N$ will be the resultant of the vibrations transmitted towards $N$ by all the points of the pupil plane.
\newline
The equi-phase wave surfaces which reach the pupil plane are not planes if the radiation has been disturbed between the source and the entrance pupil. But the hypotheses $H4$ and $H5$ make it possible to affirm that the pupil plane is spatially coherent, that is to say that at the time scale of the vibration periods, the relative phase shift of its different points is constant.
Consequently, to calculate the resulting vibration at the point $N$($p$,$q$) of the focal plane, it is necessary to sum the amplitudes that $N$ receives from the different points of the pupil plane. The amplitude distribution $a(p,q)$ in the focal plane then becomes

\begin{equation}
\label{eq:1009}
a(p,q) =  \int_{R^2} A(x,y) \exp{(-i2 \pi ( x p + y q )} dx dy,
\end{equation}

that is, the complex amplitude distribution in the focal plane $a(p,q)$ is the Fourier transform of the complex amplitude distribution $A(x,y)$ in the pupil plane, i.e.

\begin{equation}
\label{eq:1010}
a(p,q) =  FT[ A(x,y) ](p,q).
\end{equation}

\subsubsection{Applications of the fundamental theorem: the case of a single square aperture}

\begin{figure}[h]
\sidecaption
\centering
\includegraphics[width=12cm]{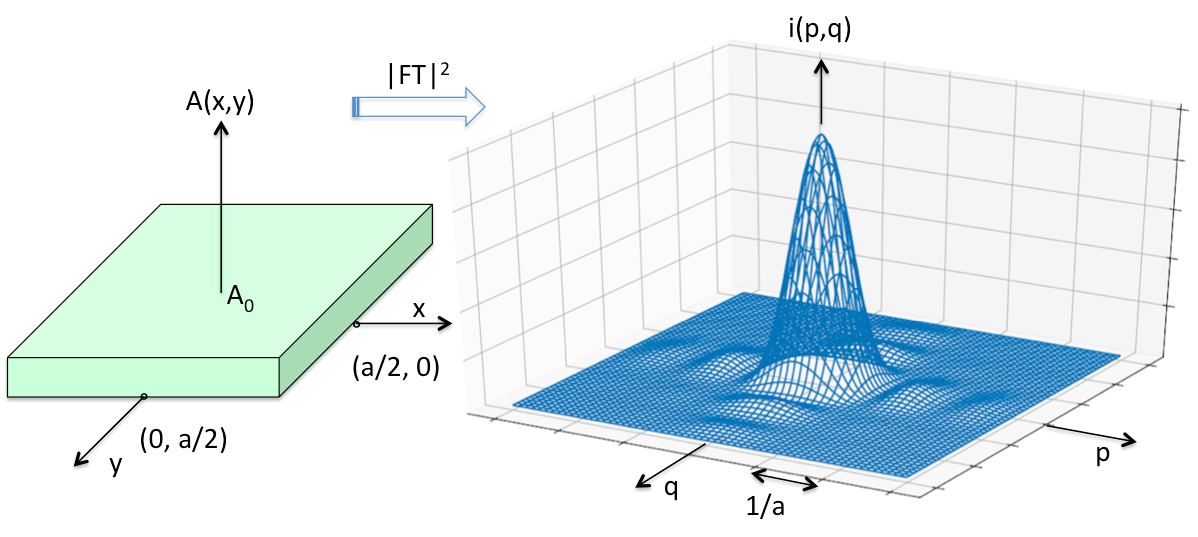}
\caption{Complex amplitude distribution $A(x,y)$ in the plane of a single square aperture (left) and resulting response function in intensity $i(p,q)$ (right).} 
\label{Fig_34_final}   
\end{figure}

Considering first the case of a single square aperture as depicted in Fig.~\ref{Fig_34_final} (left) and a point-like source perfectly located at zenith, i.e. the plane wavefronts arrive parallel to the aperture with a constant and real amplitude $A(x,y) = A_0$, we find that the calculation of the amplitude in the focal plane is very simple

\begin{equation}
\label{eq:73}
a(p,q) = A_0 \, FT[\Pi (x/a) ~ \Pi (y/a)](p,q).
\end{equation}

Making use of the separation of the variables $x, y$ and of the relation (\ref{eq:62}), Eq.~(\ref{eq:71}) successively transforms into 

\begin{equation}
\label{eq:74}
a(p,q) = A_0 \, FT[\Pi (x/a)](p) ~ FT[\Pi (y/a)](q),
\end{equation}

\begin{equation}
\label{eq:75}
a(p,q) = A_0 \, a^2 \, \frac{sin(\pi a p)}{\pi a p} \frac{sin(\pi a q)}{\pi a q}.
\end{equation}

This is the impulse response, in amplitude, for a square pupil and in the absence of any external disturbance. Adopting the definition (\ref{eq:12}) for the intensity of the vibrations, we find that (see Fig.~\ref{Fig_34_final}, at right)

\begin{equation}
\label{eq:76}
i(p,q) = a(p,q) \, a^\ast(p,q) = |a(p,q)|^2 = i_0 \, a^4 \, {[\frac{sin(\pi a p)}{\pi a p}]}^2 {[\frac{sin(\pi a q)}{\pi a q}]}^2. 
\end{equation}

Defining the angular resolution $\Phi$ of an optical system as being the angular width of the response function in intensity inside the first minima, we obtain for the values of $\pi p a = \pm \pi$ (resp.  $\pi q a = \pm \pi$), i.e. $p = \pm 1/a$ (resp. $q = \pm 1/a$) and with the definition (\ref{eq:72}) for $p, q$ 

\begin{equation}
\label{eq:77}
\frac{x'}{\lambda f} = \pm \frac{1}{a} \quad {\rm (resp.} \quad \frac{y'}{\lambda f} = \pm \frac{1}{a} {\rm )},
\end{equation}

\begin{equation}
\label{eq:78}
\Phi = \frac{\Delta x'}{f} = \frac{\Delta y'}{f} = \frac{2 \lambda}{a}.
\end{equation}

The angular resolution is thus inversely proportional to the size a of the square aperture, and proportional to the wavelength $\lambda$. Working at short wavelength with a big size aperture thus confers a better angular resolution.   \\

   Up to now, we have considered that the source $S$, assumed to be point-like and located at an infinite distance from the optical system, was on the optical axis of the instrument. Suppose now that it is slightly moved away from the zenith direction by a small angle. Let ($b/f$, $c/f$, $1$) be the unit vector representing the new direction of the source, the previous one being ($0$, $0$, $1$). The plane wavefront falling on the square aperture will not have anymore a constant amplitude $A_0$ because each point of the pupil touched by such a wavefront will experience a phase shift given by the angle

\begin{equation}
\label{eq:79}
\Psi = \frac{2\pi \delta}{\lambda} = \frac{2\pi (xb/f+yc/f)}{\lambda}
\end{equation}

and consequently the correct expression of the complex amplitude $A(x,y)$ to be inserted in Eq.~(\ref{eq:71}) becomes

\begin{equation}
\label{eq:80}
A(x,y) = A_0 \, \Pi (x/a)  \Pi (y/a) \, \exp[{\frac{i2\pi(xb/f+yc/f)}{\lambda}]}.
\end{equation}

Proceeding as previously, we easily find that 

\begin{equation}
\label{eq:81}
a(p,q) = A_0 \, FT[\Pi (\frac{x}{a})](p-\frac{b}{\lambda f}) ~ FT[\Pi (\frac{y}{a})](q-\frac{c}{\lambda f})
\end{equation}

and finally

\begin{equation}
\label{eq:82}
i(p,q) = a(p,q) \, a^\ast(p,q) = |a(p,q)|^2 = i_0 \, a^4 \, {\left[\frac{sin(\pi a (p-\frac{b}{\lambda f}))}{\pi a (p-\frac{b}{\lambda f})}\right]}^2 {\left[\frac{sin(\pi a (q-\frac{c}{\lambda f}))}{\pi a (q-\frac{c}{\lambda f})}\right]}^2 .
\end{equation}

The resulting intensity response function in the focal plane is nearly the same as the one previously calculated for the case $b=0, c=0$. It is being merely translated by a linear offset $(b,c)$ in the $x',y'$ focal plane and implies the invariance of the response function for a reference star that is being slightly offset from the optical axis of the system. 

\subsubsection{Applications of the fundamental theorem: the case of a circular aperture}

\begin{figure}[h]
\sidecaption
\centering
\includegraphics[width=12cm]{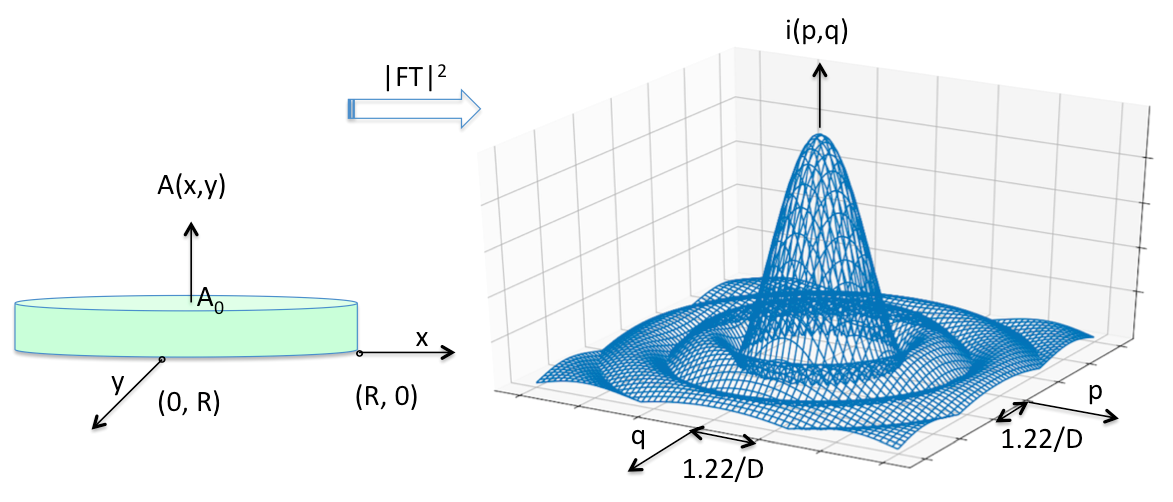}
\caption{The Airy disk: complex amplitude distribution $A(\rho, \varphi) = A_0$ in the plane of a circular aperture (left) and the resulting response function in intensity $i(\rho')$ (right).} 
\label{Fig_35_final}   
\end{figure}

Considering now a circular aperture with radius $R$, the complex amplitude $A(x,y)$ in the pupil plane may be represented as a circular symmetric distribution, i.e. $A(\rho, \varphi) = A_0$ for $\rho < R,  \varphi \in [0, 2\pi]$ and $A(\rho, \varphi) = 0$ for $\rho > R$ (see Figure \ref{Fig_35_final}, at left). 
   We naturally expect the distribution of the complex amplitude in the focal plane to be also circular symmetric, i.e.

\begin{equation}
\label{eq:83}
a(\rho') = FT[A(\rho,\varphi)](\rho').
\end{equation}

It is here interesting to note that performing the above Fourier transform is quite alike deriving the expression of the visibility of a $2{-}D$ uniform circular disk star which angular diameter is $\theta_{UD}$ (see the last application in Section 4.5). We may just make use of the result (\ref{eq:69}) with appropriate changes of the corresponding variables. We easily find that 

\begin{equation}
\label{eq:84}
a(\rho') = A_0 \pi R^2 [2 J_1(\frac{2 \pi R \rho' / (\lambda f)}{2 \pi R \rho' / (\lambda f)})].
\end{equation}

The resulting intensity response function is thus given by 

\begin{equation}
\label{eq:85}
i(\rho') = a(\rho')^2 = A_0^2 (\pi R^2)^2 [2 J_1(\frac{2 \pi R \rho' / (\lambda f)}{2 \pi R \rho' / (\lambda f)})]^2.
\end{equation}

This is the very expression of the famous Airy disk (see Fig.~\ref{Fig_35_final}, at right).

Knowing that the first order Bessel function $J_1(x) = 0$ for $x \sim 3.96$, it is easy to deduce that the angular resolution $\Phi$ of a telescope equipped with a circular objective which diameter is $D = 2R$ is given by

\begin{equation}
\label{eq:86}
\Phi = \frac{\Delta \rho'}{f} = \frac{2.44 \thinspace \lambda}{D}.
\end{equation}

\subsubsection{Applications of the fundamental theorem: the two telescope interferometer}

\begin{figure}[h]
\sidecaption
\centering
\includegraphics[width=12cm]{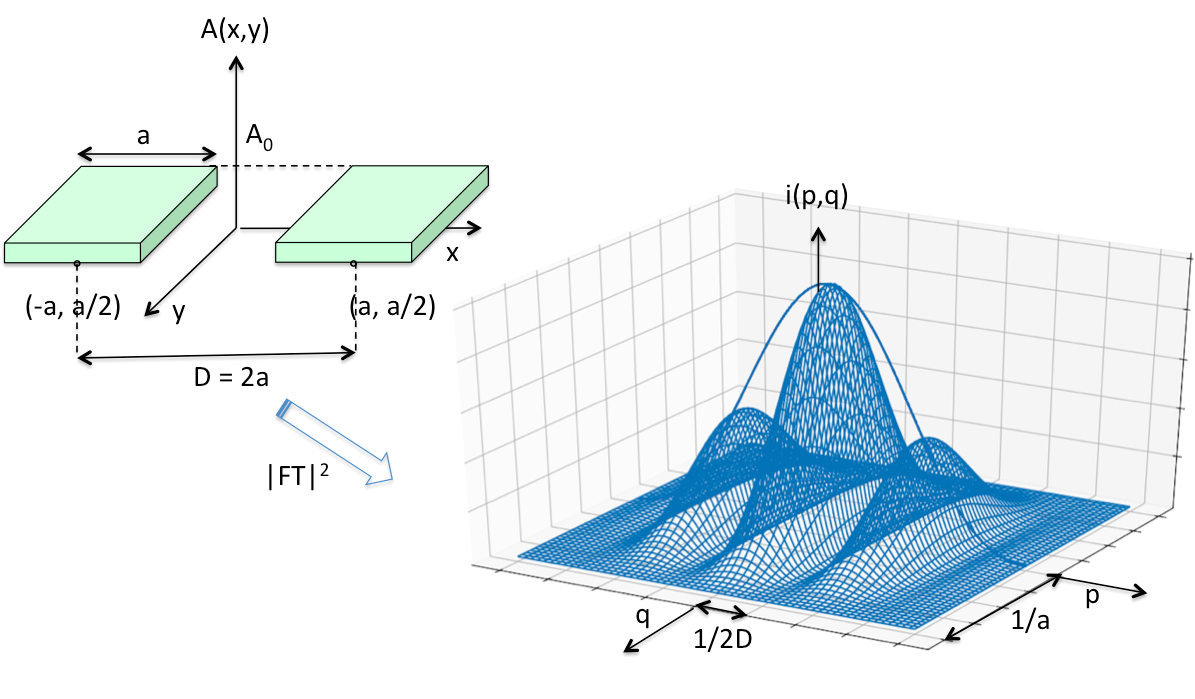}
\caption{The two telescope interferometer:  distribution of the complex amplitude for the case of two square apertures (upper left) and the corresponding impulse response function (lower right).} 
\label{Fig_36_final}   
\end{figure}

Figure \ref{Fig_36_final} (upper left) illustrates the principle of optically coupling two telescopes. Such a system is equivalent to a huge telescope in front of which would have been placed a screen pierced with two openings corresponding to the entrance pupils of the two telescopes. The pupil function $A(x, y)$ of this system is shown in that same Figure for the case of two square apertures.

Let us now calculate the impulse response function $a(p, q)$ of such a system. Representing the distribution of the complex amplitude over each of the individual square  apertures by means of the function $A_0(x,y)$ and assuming that the distance between their optical axes is D, we find that 

\begin{equation}
\label{eq:87}
a(p,q) = FT[A_0(x+D/2,y)+A_0(x-D/2,y)](p,q).
\end{equation}

Making use of the relation (\ref{eq:59}), the previous equation reduces to
\begin{equation}
\label{eq:88}
a(p,q) = [\exp({i \pi p D})+\exp({-i \pi p D})]\, FT[A_0(x,y)](p,q),
\end{equation}

\begin{equation}
\label{eq:89}
a(p,q) = 2 \, cos( \pi p D) \, FT[A_0(x,y)](p,q)
\end{equation}

and finally

\begin{equation}
\label{eq:90}
i(p,q) = a(p,q) a^{\ast}(p,q) = |a(p,q)|^2 = 4 \,cos^2( \pi p D) \, \{FT[A_0(x,y)](p,q)\}^2.
\end{equation}

Particularizing this intensity distribution to the case of two square apertures, or circular apertures, and making use of relations  (\ref{eq:76}) or (\ref{eq:85})  leads to the respective results

\begin{equation}
\label{eq:91}
i(p,q) = A_0^2 \, (2 a^2)^2 \,[\frac{sin(\pi p a)}{\pi p a}]^2 \,[\frac{sin(\pi q a)}{\pi q a}]^2 \, cos^2( \pi p D)
\end{equation}

or

\begin{equation}
\label{eq:92}
i(p,\rho') = A_0^2 (2 \,\pi R^2)^2 [\frac{2 J_1(2 \pi R \rho' / (\lambda f))}{2 \pi R \rho' / (\lambda f)}]^2 \, cos^2( \pi p D).
\end{equation}

Figure \ref{Fig_36_final} (lower right) illustrates the response function for the former case. We see that the impulse response of each individual telescope is modulated by the $\cos(2\pi pD)$ function and that the resulting impulse response function shows consequently a more detailed structure along the $p$ axis, leading to a significantly improved angular resolution $\Phi$ along that direction. The angular width $\Phi$ of the bright central fringe is equal to the angular width separating the two minima located on its two sides. We thus find successively

\begin{equation}
\label{eq:93}
\pi p D = \pm \frac{\pi}{2},
\end{equation}

\begin{equation}
\label{eq:94}
p  = \pm \frac{1}{2D}, 
\end{equation}

\begin{equation}
\label{eq:95}
\Delta p  = \frac{1}{D}
\end{equation}

and making use of relation (\ref{eq:72})
 
\begin{equation}
\label{eq:96}
\Phi  = \frac{\Delta x'}{f} = \frac{\lambda}{D}. 
\end{equation}

The angular resolution of the interferometer along the direction joining the two telescopes is approximately equivalent to that of a single dish telescope which diameter is equal to the baseline $D$ separating them, and not any longer to the diameter of each single telescope (see Eqs.~(\ref{eq:78}) or (\ref{eq:86})). 

\subsubsection{Other types of beam recombination}

\begin{figure}[h]
\centering
\includegraphics[width=8cm]{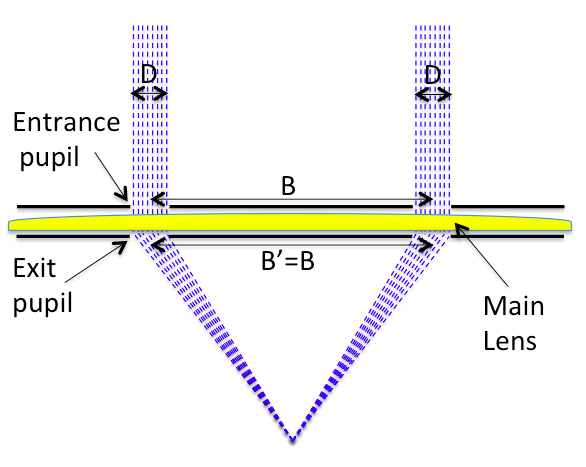}
\caption{ The two beams of light rays, represented with blue dashed lines, collected by the two entrance pupil apertures are separated by a baseline $B$ which is identical to the baseline $B'$ between the two apertures in the exit pupil plane of the main converging lens.} 
\label{Fig_1021_final}   
\end{figure}

When establishing the expression for the response function of an interferometer composed of two single square or circular apertures (see Section 6.1.3, Eqs.~(\ref{eq:91}-\ref{eq:92})), we implicitly assumed that the exit pupil perfectly matched the entrance pupil (see Figs.~\ref{Fig_14_final}, \ref{Fig_15_final}, \ref{Fig_18_final} and \ref{Fig_1021_final}).  The baseline $B$ between the two entrance pupil apertures was indeed equal to the baseline $B'$ between the two exit pupil apertures. 

\begin{figure}[h]
\centering
\includegraphics[width=8cm]{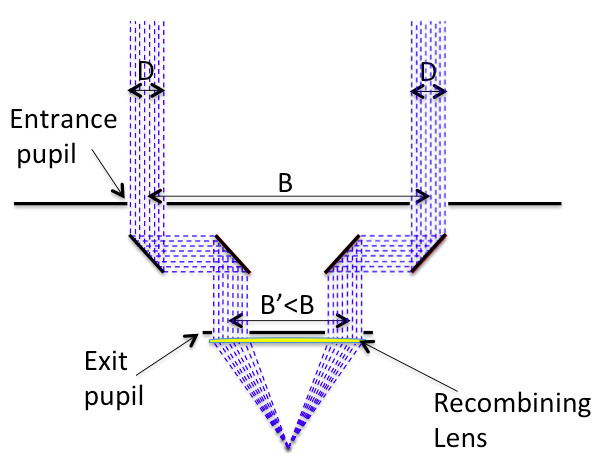}
\caption{ Sketch of the Michelson Stellar Interferometer. The baseline $B$ between the two entrance pupil apertures is much larger than the baseline $B'$ between the two apertures in front of the recombining lens. The $45^{\circ}$ inclined black lines symbolize reflective plane mirrors. In the case of the Michelson-Pease experiment, these four mirrors were set on a 7m beam just above the 2.5m Wilson telescope (see Fig. ~\ref{Fig_19_final}).} 
\label{Fig_1022_final}   
\end{figure}

This type of recombination of the two beams is referred to as the Fizeau-type or homothetic one. As we have seen in Section 3.4, Michelson and Pease have used another type of beam recombination, known as the Michelson Stellar Interferometer or still, the densified recombination type (see Fig. ~\ref{Fig_1022_final}). 

\begin{figure}[h]
\centering
\includegraphics[width=8cm]{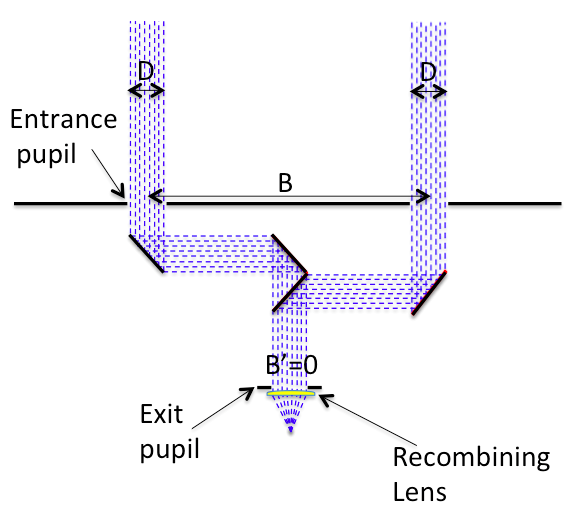}
\caption{ The Michelson Interferometer. In this case, the beam recombination is co-axial corresponding to the exit pupil baseline $B'=0$.} 
\label{Fig_1023_final}   
\end{figure}

When the two exit beams are being superimposed, resulting in the baseline $B' = 0$, the recombination is referred to as being co-axial, or the Michelson Interferometer type (see Fig.~\ref{Fig_1023_final}).  

\begin{figure}[h]
\centering
\includegraphics[width=8cm]{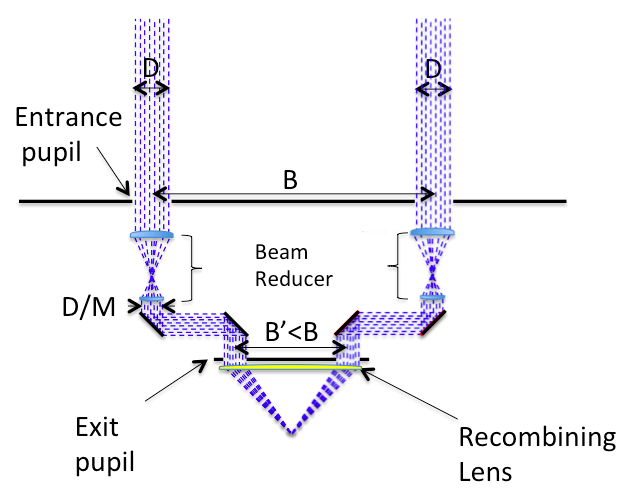}
\caption{ General case of beam recombination. The two beams of parallel light rays from a distant celestial source are first collected by two unit telescopes having a diameter $D$.  The beams are then compressed by a magnification factor M. They are subsequently relayed by a system of plane mirrors to the exit pupil of the recombining lens. At that stage, their separation (baseline) is $B' < B$. } 
\label{Fig_1024_final}   
\end{figure}

A more general model of beam recombination, that includes the three previously described ones, is illustrated in Fig.~\ref{Fig_1024_final}. Two main collectors receive the light beams from a distant celestial source. While passing through the beam reducers, the beams are compressed by a magnification factor M, corresponding to the ratio between the focal lengths of the two lenses of the focal reducers. The two compressed beams are then relayed by means of a set of 4 mirrors, just like in the Michelson Stellar Interferometer. Before entering the exit pupil of the recombining lens, their separation or baseline is $B' < B$. To calculate the response function of such an interferometer, we just need to apply the fundamental theorem to this secondary Fizeau-type interferometer with a baseline $B'$, taking into account the correct expression for the distribution of the complex amplitude of the electric field over the two exit pupil apertures. 

\begin{figure}[h]
\centering
\includegraphics[width=8cm]{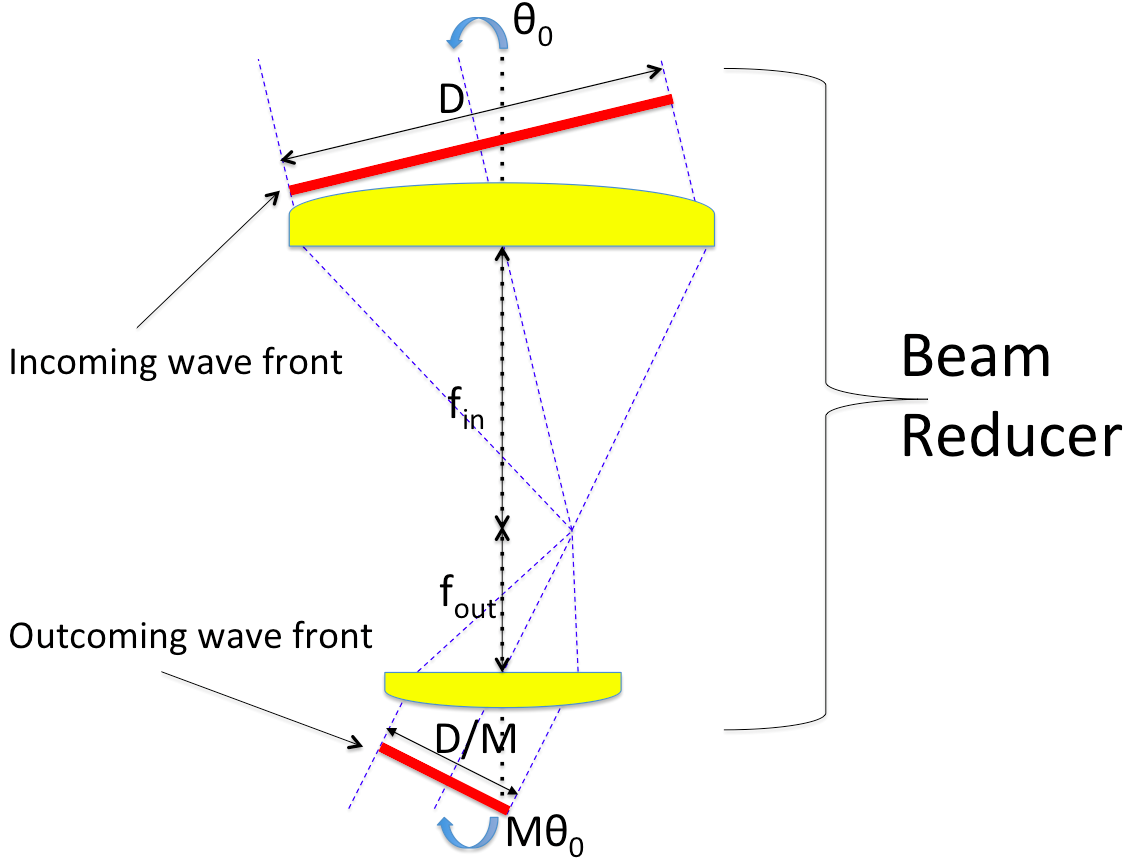}
\caption{ Propagation of an incoming plane wave from a distant celestial object with an inclination angle $\theta_0$ through a beam reducer. The beam size is being reduced by the magnification factor $M = f_{in} / f_{out}$ while the outcoming direction of the beam has changed into $M \theta_0$.  } 
\label{Fig_1025_final}   
\end{figure}

\begin{figure}[h]
\centering
\includegraphics[width=8cm]{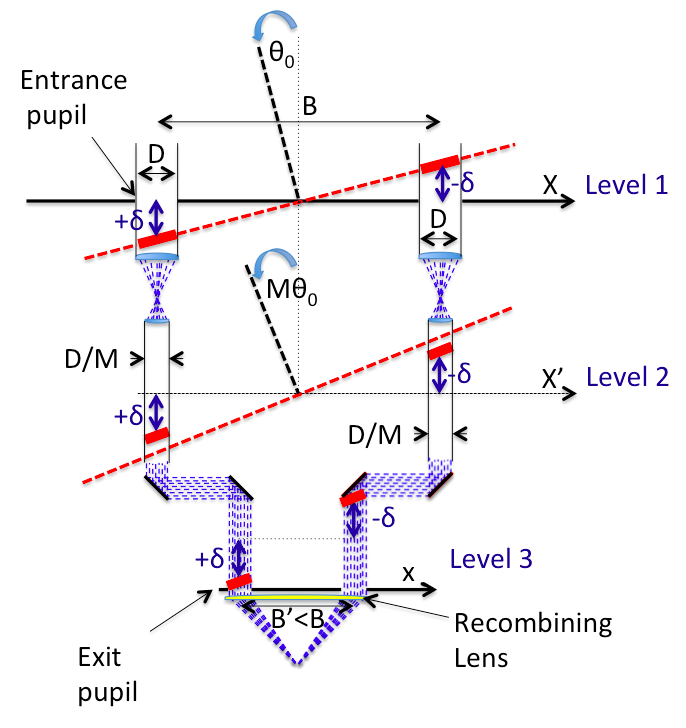}
\caption{ Propagation of an incoming plane wave from a distant celestial object with an inclination angle $\theta_0$ through two beam reducers (Level 1 - Level 2). When arriving in the exit pupil plane (Level 3), the delays $\pm \delta$ of the plane waves near the centres of the two apertures are the same but their inclination is now $M \theta_0$.} 
\label{Fig_1026_final}   
\end{figure}

Considering a point-like celestial source emitting a plane wave making an angle $\theta_0$ with respect to the line joining the two telescopes, the angle between the outcoming beam  - compressed in size by the magnification factor $M (= f_{in} / f_{out})$ - and the main axis of the optical system is $M \theta_0$ (since $\sin[{\theta_0}]  \simeq \theta_0$, given that $\theta_0 << 1$, see Fig. ~\ref{Fig_1025_final}). The resulting complex amplitude in the focal plane of the recombining lens is along the $p$ direction, i.e. along the line joining the two exit pupil apertures (see Fig. ~\ref{Fig_1026_final} and Eq.~(\ref{eq:70}))
\begin{equation}~
\label{eq:120}
a(p) = FT[A_1(x)](p) + FT[A_2(x)](p)      
\end{equation}

where $A_1(x)$ and $A_2(x)$ represent the distribution of the complex amplitude in the two exit pupil apertures along the $x$ axis.

We subsequently find that 
\begin{equation}
\label{eq:121}
\begin{split}
FT[A_1(x)](p) = \int_{-(B'+D/M)/2 }^{-(B'-D/M)/2} M \exp[{-2i \pi(p x)}] 
\exp[{2i \pi M \sin[{\theta_0}] (x - (B/M-B')/2) / \lambda }] dx,  \\ 
FT[A_2(x)](p) = \exp[{2i \pi(d / \lambda)}] \int_{(B'-D/M)/2 }^{(B'+D/M)/2} M \exp[{-2i \pi(p x)}] \exp[{2i \pi M \sin[{\theta_0}] (x + (B/M-B')/2) / \lambda }] dx.
\end{split}
\end{equation}

In this expression, we have taken into account the fact that most of existing interferometers are equipped with a delay line and we have assumed here that an extra length $d$ affects the path of the second beam. This explains the origin of the factor $\exp[{2i \pi(d / \lambda)}]$  in the expression of  $ FT[A_2(x)](p)$.  The limits of integration are straightforward to establish (see Fig. ~\ref{Fig_1026_final}, Level 3). The presence of the factor $M$ merely accounts for the fact that when a beam is compressed, its constant amplitude is being multiplied by M (and the intensity $i(p)$ by $M^2$ in order to preserve energy conservation). The factor $\exp[{-2i \pi(p x)}]$  merely accounts for the pupil-to-image relationship from Fourier optics (cf. the fundamental theorem). Since for the case of a co-phased array, the path differences affecting the arrival of the plane waves at the centres of the two apertures at Level 1 in Fig.~\ref{Fig_1026_final} are $+\delta$ and $-\delta$ ($= \pm (B/2) \sin[{\theta_0}] \simeq \pm (B/2) \theta_0$), the latter remain unaffected when reaching the centres of the two apertures in the exit pupil plane (Level 3). Nevertheless, their relative inclination has changed from $\theta_0$ to $M \theta_0$. Therefore, we easily understand the origin of the two factors $\exp[{2i \pi M \sin[{\theta_0}] (x - (B/M-B')/2) / \lambda }]$ and $\exp[{2i \pi M \sin[{\theta_0}] (x + (B/M-B')/2) / \lambda }]$ appearing in the two previous equations. After several successful changes of variables (see Appendix), Eq. (\ref{eq:120}) reduces to   

\begin{equation}
\label{eq:122}
a(p) = 2D \exp[{i \pi(d / \lambda)}] \,\frac{\sin[{(\pi D / M) (p - M \sin[{\theta_0}] / \lambda)}]} {(\pi D / M) (p - M \sin[{\theta_0}] / \lambda)} \, \cos[{\pi (B' p + (d - B \sin[{\theta_0}]) / \lambda})].  
\end{equation}
The corresponding expression for the intensity $i(p) = |a(p)|^2$ becomes
 
\begin{equation}
\label{eq:123}
i(p) = 4D^2 \,[\frac{\sin[{(\pi D / M) (p - M \sin[{\theta_0}] / \lambda)}]} {(\pi D / M) (p - M \sin[{\theta_0}] / \lambda)}]^2 \, [\cos[{\pi (B' p + (d - B \sin[{\theta_0}]) / \lambda})]]^2.  
\end{equation}

The previous equations describe the response function of any interferometer having its entrance and exit baselines such as $0 \leq B' \leq B$.

In the absence of an internal delay $d$, the previous expression for $i(p)$ can be rewritten as

\begin{equation}
\label{eq:124}
i(p) = 4D^2 \,[\frac{\sin[{(\pi D / M) (p - M \sin[{\theta_0}] / \lambda)}]} {(\pi D / M) (p - M \sin[{\theta_0}] / \lambda)}]^2 \, [\cos[{\pi B' (p  - B \sin[{\theta_0}] / (B' \lambda})]]^2.  
\end{equation}

Some nice features become outstanding: we first notice that the width of the envelope function is governed by the factor $\pi D / M$ which is related to the size of the beam after compression. The angular separation of the fringes ($\lambda / B'$) is essentially determined by the exit pupil baseline $B'$. It does neither depend on the main baseline $B$ nor on the magnification (or beam compression) $M$. This last equation also reveals that for the response function to be field invariant, we must have $M = B / B'$. In that case, the centre of the main envelope (cf. Airy disk for the case of a circular aperture) will always coincide with the central fringe peak, whatever the position ($\theta_0$) of the source in the field of view. 

Let us now consider the case of Fizeau-type interferometry for which we have $d = 0$ (no delay line is being used) and in addition $M = 1$, $B' = B$, also $\sin[{\theta_0}] \simeq \theta_0$, Eq.~(\ref{eq:124}) then reduces to

\begin{equation}
\label{eq:125}
i(p) = 4D^2 \,[\frac{\sin[{(\pi D ) (p - \theta_0 / \lambda)}]} {(\pi D ) (p -  \theta_0 / \lambda)}]^2 \, [\cos[{\pi (B (p - \theta_0 / \lambda)})]]^2.  
\end{equation}

Posing $\theta_0 = b/f $ in the latter equation, we simply recover the result previously established for the case of Fizeau interferometry (see Eqs.~(\ref{eq:82}) and (\ref{eq:91})). We also note here that the response function of a Fizeau-type interferometer is field invariant.

Finally, the response function of a co-axial interferometer is easily derived by inserting the value $B' = 0$ in Eq.~(\ref{eq:123}):

\begin{equation}
\label{eq:126}
i(p) = 4D^2 \,[\frac{\sin[{(\pi D / M) (p - M \theta_0 / \lambda)}]} {(\pi D / M) (p - M \theta_0 / \lambda)}]^2 \, [\cos[{\pi (d - B \theta_0) / \lambda}]]^2.  
\end{equation}

We note here that the $cos$ factor is only a function of $d$ and $\theta_0$, and not any longer of  $p$.

In conclusion, we have established in this section a very general expression (see Eq.~(\ref{eq:123})) for the response function of an interferometer composed of two similar apertures separated by a baseline $B$ and which beams have been compressed by a magnification factor $M$. In the exit pupil plane, the new baseline between the two beams is $B'$ ($0 < B' < B$) such that the fringe separation is essentially governed by the latter term.

\subsection{The convolution theorem}

The fundamental theorem has allowed us to take into account the finite size of the apertures of an optical system instead of considering that the apertures are made of pinholes. However, we have considered that the source is point-like. To treat the case of an extended source, we shall make use of the convolution theorem.  

The convolution theorem states that the convolution of two functions $f(x)$ and $g(x)$ is given by the following expression

\begin{equation}
\label{eq:97}
f(x) \ast g(x)=(f \ast g)(x)=\int_{R}f(x-t)g(t)dt. 
\end{equation}

\begin{figure}[h]
\sidecaption
\centering
\includegraphics[width=13cm]{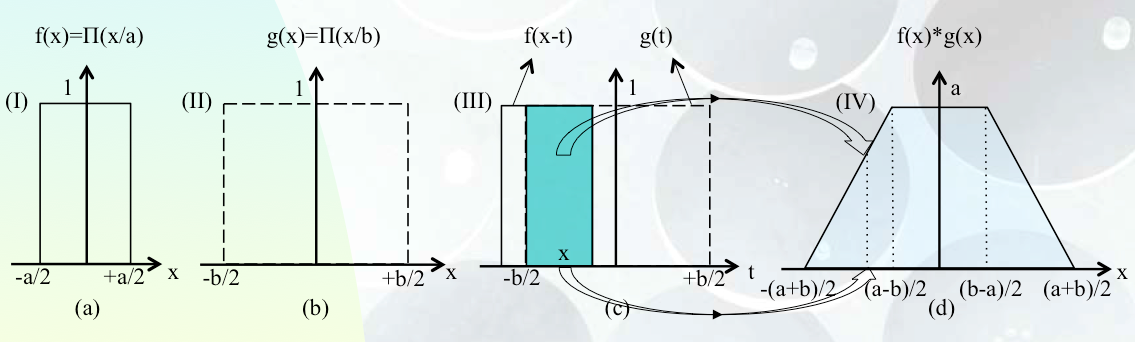}
\caption{Convolution product of two 1-D rectangular functions. (a) $f(x)$, (b) $g(x)$, (c) $g(t)$ and $f(x-t)$. The dashed area represents the integral of the product of $f(x-t)$ and $g(t)$ for the given $x$ offset, (d) $f(x)\ast g(x)$ = $(f \ast g)(x)$ represents the previous integral as a function of $x$.} 
\label{Fig_37_final}   
\end{figure}

Figure \ref{Fig_37_final} illustrates such a convolution product for the case of two rectangular functions $f(x) = \Pi(x/a)$ and $g(x) = \Pi(x/b)$ having the widths $a$ and $b$, respectively.
 
\begin{figure}[h]
\sidecaption
\centering
\includegraphics[width=9cm]{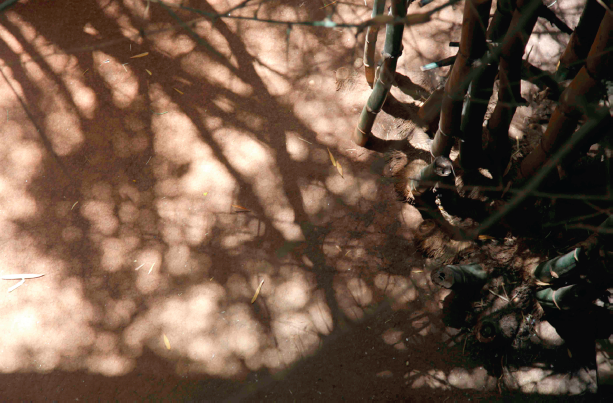}
\caption{Projected images of the Sun on the ground actually produced through small holes in the foliage of trees (bamboo trees at IUCAA, Pune, India, June 2016). These images actually result from the convolution of the intrinsic Sun intensity distribution and the shapes of the holes in the trees.} 
\label{Fig_38_final}   
\end{figure}

Every day when the Sun is shining, it is possible to see nice illustrations of the convolution product while looking at the projected images of the Sun on the ground which are actually produced through small holes in the foliage of the trees (see the illustration in Fig.~\ref{Fig_38_final}). It is a good exercise to establish the relation existing between the observed surface brightness of those Sun images, the shape of the holes in the foliage of the trees, their distance from the ground and the intrinsic surface brightness distribution of the Sun. 

\subsubsection{Application to the case of the two telescope interferometer}

We have previously seen that for the case of a point-like source having an intrinsic surface brightness distribution $O(p,q) = \delta(p) \delta(q)$, there results the formation of an image $e(p,q)$ in the focal plane which is the impulse response $e(p,q) = i(p,q) = |a(p,q)|^2$ of the optical instrument (see Eqs.~(\ref{eq:76}), (\ref{eq:85}), (\ref{eq:91}), (\ref{eq:92}) for the case of a single square aperture, a single circular aperture, an interferometer composed of two square or circular apertures, respectively).  Considering now an extended source represented by its intrinsic surface brightness distribution $O(p,q)$, application of the convolution theorem in two dimensions directly leads to the expression of its brightness distribution $e(p,q)$ in the focal plane of the optical system

\begin{equation}
\label{eq:98}
e(p,q) = O(p,q) \ast |a(p,q)|^2
\end{equation}

or more explicitly

\begin{equation}
\label{eq:99}
e(p,q) = \int_{R^2} O(r,s)  |a(p-r,q-s)|^2 dr ds.
\end{equation}

Since the Fourier transform of the convolution product of two functions is equal to the product of their Fourier transforms, we find that   

\begin{equation}
\label{eq:100}
FT[e(p,q)] = FT[O(p,q)] \, FT[|a(p,q)|^2]
\end{equation}

and also that the inverse Fourier transform of $FT[O(p,q)]$ leads to the result

\begin{equation}
\label{eq:101}
O(p,q) = FT^{-1}[FT[O(p,q)]] = FT^{-1}[\frac{FT[e(p,q)]}{FT[|a(p,q)|^2]}],
\end{equation}

namely, that it should be possible to recover interesting information on the intrinsic surface brightness distribution of the source $O(p,q)$ at high angular resolution provided that we get sufficient information at high frequencies in the $u,v$ plane on the object $FT[e(p,q)]$ itself as well as on a reference point-like object $FT[|a(p,q)|^2]$.

\subsubsection{Interferometric observations of a circular symmetric source}

Considering the case of a symmetric source around the $Y$ axis observed by means of an interferometer composed of two square apertures which size of their sides is $d$ separated along the $X$ axis by the baseline $D$, we find by means of Eqs.~(\ref{eq:76}), (\ref{eq:91}) and (\ref{eq:98}) that   

\begin{equation}
\label{eq:102}
e(p)=2d^2 \left(\dfrac{\sin(\pi pd)}{\pi pd}\right)^2
\left[O(p)\ast \cos^2
(\pi pD)\right].
\end{equation}

Making use of the relation $\cos(2x) = 2 \cos^{2}(x) - 1$, Eq.~(\ref{eq:102}) reduces to

\begin{equation}
\label{eq:103}
e(p)=2d^2 \left(\dfrac{\sin(\pi pd)}{\pi pd}\right)^2
\left[\dfrac{1}{2} \int_R O(p)dp+\dfrac{1}{2}\, O(p)\ast \cos
(2\pi pD)\right].
\end{equation}

Since the function $O(p)$ is real, the previous relation may rewritten in the form

\begin{equation}
\label{eq:104}
e(p)= A \left [ B + \dfrac{1}{2} Re [ O(p) \ast \exp({i 2 \pi pD}) ] \right ]
\end{equation}

where 

\begin{equation}
\label{eq:105}
A = 2d^2 \left(\dfrac{\sin(\pi pd)}{\pi pd}\right)^2 \quad {\rm and} \quad B = \dfrac{1}{2} \int_R O(p)dp.
\end{equation}

Given the definition of the convolution product (cf. Eq.~(\ref{eq:97})), relation (\ref{eq:104}) can be rewritten as 

\begin{equation}
\label{eq:106}
e(p) = A \left [ B + \dfrac{1}{2} Re [ \int_R O(r) \, \exp({i 2 \pi (p-r) D}) dr] \right ],
\end{equation}

or 

\begin{equation}
\label{eq:107}
e(p) = A \left [ B + \dfrac{1}{2} cos(2 \pi p D) \, FT[O(r)](D)  \right ],
\end{equation}

because $O(p)$ being real and even, its Fourier transform is also real. The visibility of the fringes being defined by (see Eq.~(\ref{eq:21})) 

\begin{equation}
\label{eq:108}
v=|\gamma_{12}(D)|=\left(\dfrac{e_{\max}-e_{\min}}{e_{\max}+e_{\min}}\right),
\end{equation}

we obtain

\begin{equation}
\label{eq:109}
v=|\gamma_{12}(D)|=FT[\frac{O(r)}{2B}](D) = FT[\frac{O(r)}{\int O(p) dp}](D).
\end{equation}

We have thus recovered the important result (see Eq. (\ref{eq:48}), i.e. the Zernicke-van Cittert Theorem), first established for the case of two point-like apertures, according to which the visibility of the fringes is the Fourier transform of the normalized intensity distribution of the source. This result can be generalized to the case of a source that is not symmetric.

\subsection{The Wiener-Khinchin theorem}

Finally, the Wiener-Khinchin theorem allows one to easily figure out what is the space frequency content of the point spread function for a given entrance pupil of an optical instrument. We may then directly find out which information is recoverable in terms of space frequency when observing an extended source.  

The Wiener-Khinchin theorem merely states that the Fourier transform of the response function of an optical system, i.e. the Fourier transform of the Point Spread Function in our case, is given by the auto-correlation of the distribution of the complex amplitude in the pupil plane. In mathematical terms, the theorem can be expressed as follows 

\begin{equation}
\label{eq:110}
FT[|a(p,q)|^2](x,y)=FT[i(p,q)](x,y)= \int_{-\infty}^{+\infty}\int_{-\infty}^{+\infty}
A^\ast(x'+x,y'+y)A(x',y')dx'dy'.
\end{equation}

\begin{figure}[h]
\sidecaption
\centering
\includegraphics[width=9cm]{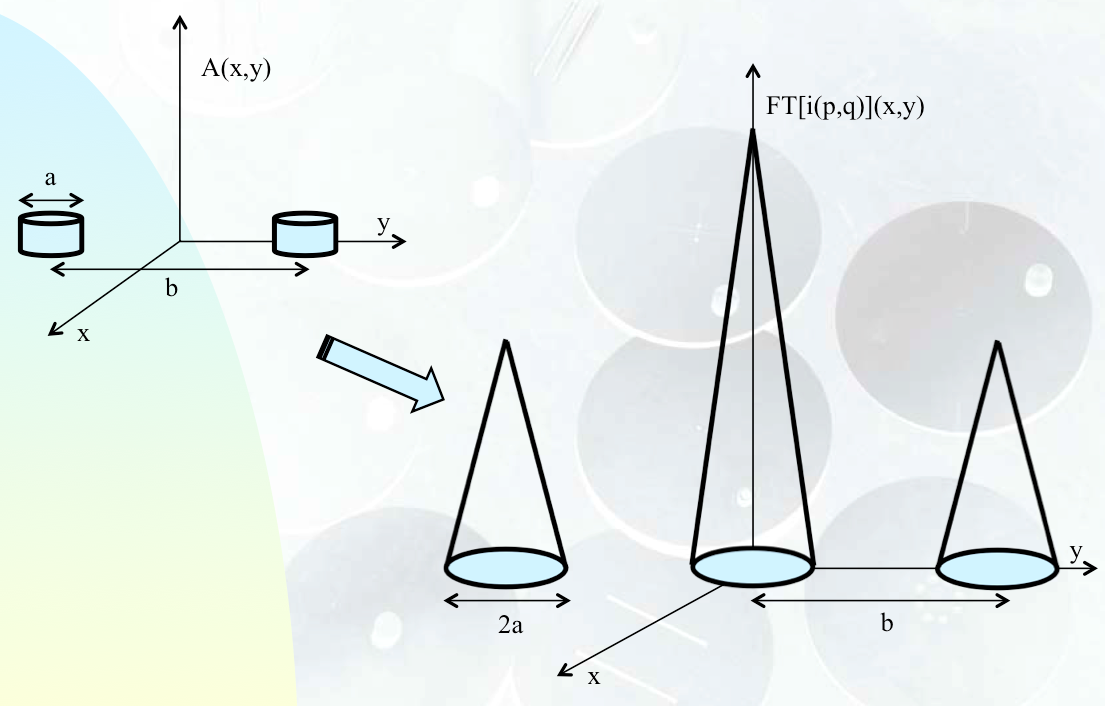}
\caption{Diagram representing the autocorrelation function versus the space frequency, for a two telescope interferometer, each having a diameter $a$, separated by the baseline $b$. } 
\label{Fig_39_final}   
\end{figure}

When establishing the expression (\ref{eq:101}), we wrote that the quantity $FT[|a(p,q)|^2]$ appearing in its denominator could be retrieved from the observation of a point-like star. The Wiener-Khinchin theorem states that it can also be retrieved from the auto-correlation of the distribution of the complex amplitude $A(x,y)$ in the pupil plane. Figure \ref{Fig_39_final} illustrates the application of this theorem to the case of an interferometer composed of two circular apertures having a diameter $a$ and separated by the baseline $b$. We see that the autocorrelation of an interferometer gives access to high space frequencies.

A simple demonstration of the Wiener-Khinchin theorem (\ref{eq:110}) is given below.

We may successively establish that

\begin{equation}
\label{eq:111}
FT[i(p,q)](x,y)=FT[|a(p,q)|^2](x,y)=FT[a^\ast(p,q) a(p,q)](x,y),
\end{equation}

\begin{equation}
\label{eq:112}
\begin{split}
FT[i(p,q)](x,y)=\int \int \exp[{-2i \pi(px+qy)}] \int \int A^\ast(x'',y'') \, \exp[{2i \pi(px''+qy'')}] \, dx''dy'' \\ \int \int A(x',y') \, \exp[{-2i \pi(px'+qy')}] \, dx'dy' dp dq,
\end{split}
\end{equation}

\begin{equation}
\label{eq:113}
\begin{split}
FT[i(p,q)](x,y)= \int \int \exp[(2i \pi \{ p[x''-(x'+x)]+q[y''-(y'+y)] \})]  \\ \int \int A^\ast(x'',y'') \, dx''dy''  \int \int A(x',y') \,  dx' dy' dp dq
\end{split}
\end{equation}

and taking into account the definition (\ref{eq:63}) of the Dirac distribution
\begin{equation}
\label{eq:114}
FT[i(p,q)](x,y)=\int \int  \int \int  \delta [x''-(x'+x)] \, \delta [y''-(y'+y)] \, A^\ast(x'',y'') \,  A(x',y') \, dx'dy' dx''dy''.
\end{equation}
We finally find that
\begin{equation}
\label{eq:115}
\begin{split}
FT[i(p,q)](x,y)=\int \int A^\ast(x'+x,y'+y) \,  A(x',y') \, dx'dy' dx''dy''  = \\ \int_{-\infty}^{+\infty} \int_{-\infty}^{+\infty} A^\ast(x'+x,y'+y) \,  A(x',y') \, dx'dy' dx''dy'',
\end{split}
\end{equation}
i.e. the quoted result, namely that the Fourier transform of the impulse response function of an optical system can be represented by the autocorrelation of the distribution of the complex amplitude $A(x,y)$ in the pupil plane.\\
\vspace{0.3cm}

These lecture notes are based upon lectures on the same subject delivered by the author in French at the Li\`{e}ge University during the past ten years (see [1]). To get deeper into the field of interferometry, we highly recommend the following books: [2], [3], [4], [5].\\
\vspace{0.3cm}

Finally, I wish to thank the organizers of the 2017 Evry Schatzman School (Dr. N. Nardetto, Prof. Y. Lebreton, Dr. E. Lagadec and Dr. A. Meilland) for their invitation to give these lectures and for the warm hospitality and nice atmosphere in Roscoff during that event.\\   
\vspace{0.3cm}

\subsection{Appendix}

In this appendix, we detail the calculations leading from Eqs.~(\ref{eq:120})-(\ref{eq:121}) to Eq.~(\ref{eq:122}).

First of all, we proceed with the following change of variables in the expression of $FT[A_1(x)](p)$: $y = -x, dy = -dx$. We then replace $y$ by $x$ and $dy$ by $dx$. Putting the factor $\exp[i \pi d / \lambda]$ in evidence, the summation of $FT[A_1(x)](p)$ and $FT[A_2(x)](p)$ leads to

\begin{equation}
\label{eq:127}
\begin{split}
a(p) = \exp[{i \pi(d / \lambda)}] \{ \exp[{-i \pi(d / \lambda)}] M \exp[{-i \pi M \sin[{\theta_0}] (B/M-B') / \lambda }] \cdot \\ 
\int_{(B'-D/M)/2 }^{(B'+D/M)/2} \exp[{2i \pi x (p - M \sin[{\theta_0}] / \lambda)}] dx + \\
\exp[{i \pi(d / \lambda)}] M \exp[{i \pi M \sin[{\theta_0}] (B/M-B') / \lambda }] \cdot \\
\int_{(B'-D/M)/2 }^{(B'+D/M)/2} \exp[{-2i \pi x (p - M \sin[{\theta_0}] / \lambda)}] dx \}
\end{split}
\end{equation}

and subsequently

\begin{equation}
\label{eq:128}
\begin{split}
a(p) = M \exp[{i \pi(d / \lambda)}] \int_{(B'-D/M)/2 }^{(B'+D/M)/2} \{ \exp[{i \pi[2x (p - M \sin[{\theta_0}] / \lambda) - (d + M \sin[{\theta_0}] (B/M - B')}] + \\
\exp[{-i \pi[2x (p - M \sin[{\theta_0}] / \lambda) - (d + M \sin[{\theta_0}] (B/M - B')}] dx \}, 
\end{split}
\end{equation}

\begin{equation}
\label{eq:129}
\begin{split}
a(p) = 2 M \exp[{i \pi(d / \lambda)}] \int_{(B'-D/M)/2 }^{(B'+D/M)/2}  \cos[{\pi[2x (p - M \sin[{\theta_0}] / \lambda) - (d + M \sin[{\theta_0}] (B/M - B')}] dx. 
\end{split}
\end{equation}

Let us now make use of the change of variables \\ 
$z = \pi[2x (p - M \sin[{\theta_0}] / \lambda) - (d + M \sin[{\theta_0}] (B/M - B')]$ such that 
$dx = dz / [2 \pi (p - M \sin[{\theta_0}] / \lambda)] $, Eq.~(\ref{eq:129}) then transforms into

\begin{equation}
\label{eq:130}
\begin{split}
a(p) = \frac{2 M \exp[{i \pi(d / \lambda)}]}{2 \pi (p - M \sin[{\theta_0}] / \lambda)} \{ \sin[{\pi \{ (B' + D/M)(p - M \sin[{\theta_0}] / \lambda) - (d + M \sin[{\theta_0}](B/M - B')) / \lambda \}}] \\ - \sin[{\pi \{ (B' - D/M)(p - M \sin[{\theta_0}] / \lambda) - (d + M \sin[{\theta_0}](B/M - B')) / \lambda \}}] \}, 
\end{split}
\end{equation}

and still

\begin{equation}
\label{eq:131}
\begin{split}
a(p) = \frac{D \exp[{i \pi(d / \lambda)}]}{\pi D [p - M \sin[{\theta_0}] / \lambda]/M} \{ \sin[\Gamma + \Lambda] - \sin[\Gamma - \Lambda]\}, \\ 
\rm{with} \thinspace \Gamma = \pi \{ B' (p - M \sin[{\theta_0}] / \lambda) - (d + M \sin[{\theta_0}](B/M - B')) / \lambda \}, \\
\rm{and} \thinspace \Lambda =  \pi D (p - M \sin[{\theta_0}] / \lambda) / M.
\end{split}
\end{equation}

Making use of the well known relation $\sin(\Gamma+\Lambda) - \sin(\Gamma-\Lambda) = 2 \cos(\Gamma) \sin(\Lambda)$, the previous equation reduces to

\begin{equation}
\label{eq:132}
\begin{split}
a(p) = 2 D \exp[{i \pi(d / \lambda)}] \frac{\sin[{\pi D [(p - M \sin[{\theta_0}] / \lambda)]/M}]}{\pi D [(p - M \sin[{\theta_0}] / \lambda)]/M} \cdot \\
\cos[\pi \{ B' (p - M \sin[{\theta_0}] / \lambda) -(d + M \sin[{\theta_0}] (B/M - B')) / \lambda  \}] 
\end{split}
\end{equation}

and since 

\begin{equation}
\label{eq:1132}
\pi \{ B' (p - M \sin[\theta_0] / \lambda) -(d + M \sin[\theta_0] (B/M - B')) / \lambda  \} = \pi \{ B' p + (d - B \sin[\theta_0]) / \lambda \},
\end{equation}

we finally obtain

\begin{equation}
\label{eq:133}
\begin{split}
a(p) = 2 D \exp[{i \pi(d / \lambda)}] \frac{\sin[{\pi D [(p - M \sin[{\theta_0}] / \lambda)]/M}]}{\pi D [(p - M \sin[{\theta_0}] / \lambda)]/M} \cdot \\
\cos[\pi \{ B' p + (d - B \sin[{\theta_0}]) / \lambda\}] 
\end{split}
\end{equation}
which is the same result as that quoted in Eq.~(\ref{eq:122}).

\vspace{2cm}

{\bf References}\\   
\vspace{0.3cm}

1. \thinspace J. Surdej, see http://www.aeos.ulg.ac.be/teaching.php (2018)\\
 
2. \thinspace H. R\'{e}boul, {\it Introduction \`{a} la th\'{e}orie de l'Observation en Astrophysique} (Masson, 1979) \\
 
3. \thinspace P. L\'{e}na, D. Rouan, F., Lebrun, F. Mignard, D., Pelat, D., {\it Observational Astrophysics} (Astronomy and Astrophysics Library, 2012) \\

4. \thinspace A. Glindemann, {\it Principles of Stellar Interferometry} (Astronomy and Astrophysics Library, 2011)\\

5. \thinspace D. Buscher, {\it Practical Optical Interferometry: Imaging at Visible and Infrared Wavelengths} (Cambridge University Press, 2015)\\

\end{document}